\documentclass{amsart}
\usepackage{epsfig}
\usepackage{graphics}
\usepackage{amsmath}
\usepackage{amssymb}

\DeclareMathOperator{\sign}{sign}
\DeclareMathOperator{\Spin}{Spin}
\DeclareMathOperator{\SU}{SU}
\DeclareMathOperator{\SO}{SO}
\DeclareMathOperator{\E}{E}

\newcommand{\f}{f}
\newcommand{\F}{F}

\newcommand{\K}{K}
\newcommand{\I}{I}
\newcommand{\V}{V}

\newtheorem{conj}{Conjecture}
\newcommand{\maps}{\colon}
\newcommand{\R}{\mathbb{R}}
\newcommand{\N}{\mathcal{N}}
\newcommand{\X}{\mathcal{X}}
\renewcommand{\H}{\mathcal{H}}

\newcommand{\riemannianKernel}[2] {\K^R_{#1}(#2)}
\newcommand{\euclideanKernel}[2] {\K^D_{#1}(#2)}
\newcommand{\lorentzianKernel}[2] {\K^L_{#1}(#2)}

\newcommand{\riemannianIntegral}[1] {\I^R(#1)}
\newcommand{\euclideanIntegral}[1] {\I^D(#1)}
\newcommand{\lorentzianIntegral}[1] {\I^L(#1)}

\newcommand{\riemannianIntegralDeg}[1] {\I^R_{\rm deg}(#1)}
\newcommand{\riemannianIntegralStat}[1] {\I^R_{\rm stat}(#1)}

\newcommand{\smult}{\lambda}

\newcommand{\revisions}[1] {}

\newcommand{\cxymatrix}[1]{\vcenter{\xymatrix{#1}}}

\input xy \xyoption{all} \xyoption{poly} % use Xy-pic
\labelmargin-{1.2pt}  % bring the labels closer

\newcommand{\thetagraph}[3]  % 2 vertices 3 edges 
{ 
\xymatrix{ *{\bullet} 
\ar@{-} 
@/^1.5pc/ 
[r] 
^{#1} 
\ar@{-} 
@/_1.5pc/ 
[r] 
_{#3} 
\ar@{-} 
[r]^{#2} 
& 
*{\bullet} 
\\} 
} 

\newcommand{\fourtheta}[4]  % 2 vertices 4 edges 
{ 
\cxymatrix{ *{\bullet} 
\ar@{-} 
@/^1.5pc/ 
[r] 
^{#1} 
\ar@{-} 
@/_1.5pc/ 
[r] 
_{#4} 
\ar@{-} 
@/^/ 
[r]^{#2} 
\ar@{-} 
@/_/ 
[r]_{#3} 
& 
*{\bullet} 
\\} 
} 

\newcommand{\TetJ}[6]{
\def\lab{\ifcase\xypolynode\or #1 \or #2 \or #3 \fi}
\begin{xy} 
\xygraph{!{<3.2pc,0pc>:}
  *{\bullet}
  !P3"A"{~><{@{{}{-}*{\bullet}}} ~>>{_{\lab}}}
  "A0" -@-_{#4} "A1"
  "A0" -@-_{#5} "A2"
  "A0" -@-^{#6} "A3"
}
\end{xy}
}

\newcommand{\TenJ}{
\def\lab{\ifcase\xypolynode\or 12 \or 23 \or 34 \or 45 \or 15 \fi}
\begin{xy} 
\xygraph{!{<4pc,0pc>:}
  !P5"A"{~><{@{{}{-}*{\bullet}}} ~>>{_{j_{\lab}}}}
  "A1" -@-_{j_{13}} "A3"
  "A2" -@-_{j_{24}} "A4"
  "A3" -@-_{j_{35}} "A5"
  "A4" -@-_{j_{14}} "A1"
  "A5" -@-_{j_{25}} "A2"
}
\end{xy}
}

\newcommand{\Ten}{%
\def\lab{\ifcase\xypolynode\or 1,2 \or 2,3 \or 3,4 \or 4,5 \or 5,1 \fi}
\begin{xy} 
\xygraph{!{<1pc,0pc>:}
  !P5"A"{~><{@{{}{-}*{\bullet}}} ~>>{_{}}}
  "A1" -@-_{} "A3"
  "A2" -@-_{} "A4"
  "A3" -@-_{} "A5"
  "A4" -@-_{} "A1"
  "A5" -@-_{} "A2"
}
\end{xy}
}

\newcommand{\TenL}{
\def\lab{\ifcase\xypolynode\or 1,2 \or 2,3 \or 3,4 \or 4,5 \or 5,1 \fi}
\begin{xy} 
\xygraph{!{<2pc,0pc>:}
  !P5"A"{~><{@{{}{-}*{\bullet}}} ~>>{_{\smult}}}
  "A1" -@-_{\smult} "A3"
  "A2" -@-_{\smult} "A4"
  "A3" -@-_{\smult} "A5"
  "A4" -@-_{\smult} "A1"
  "A5" -@-_{\smult} "A2"
}
\end{xy}
}

\begin{document}

\title{Asymptotics of $10j$ symbols}
\author{John C. Baez}
\address{Department of Mathematics\\
         University of California\\
         Riverside\\
         California 92521\\
         U.S.A.}
\email{baez@math.ucr.edu}
\author{J. Daniel Christensen}
\address{Department of Mathematics\\
         University of Western Ontario\\
         London\\
         ON N6A 5B7\\
         Canada}
\email{jdc@uwo.ca}
\author{Greg Egan}
\email{gregegan@netspace.net.au}

\date{October 20, 2002}

\begin{abstract} 
\noindent 
The Riemannian $10j$ symbols are spin networks that assign an
amplitude to each 4-simplex in the Barrett-Crane model of Riemannian
quantum gravity.  This amplitude is a function of the areas of the 10
faces of the 4-simplex, and Barrett and Williams have shown that one
contribution to its asymptotics comes from the Regge action for all
non-degenerate 4-simplices with the specified face areas.  However, we
show numerically that the dominant contribution comes from degenerate
4-simplices.  As a consequence, one can compute the asymptotics of the
Riemannian $10j$ symbols by evaluating a `degenerate spin network',
where the rotation group $\SO(4)$ is replaced by the Euclidean group of
isometries of $\R^3$.  We conjecture formulas for the asymptotics of a
large class of Riemannian and Lorentzian spin networks in terms of these
degenerate spin networks, and check these formulas in some special
cases.  Among other things, this conjecture implies that the Lorentzian
$10j$ symbols are asymptotic to $1/16$ times the Riemannian ones.
\end{abstract}

\maketitle

\section{Introduction}

In the Ponzano--Regge model of 3-dimensional Riemannian quantum
gravity~\cite{PR}, an amplitude is associated with each tetrahedron in
a triangulation of spacetime.  The amplitude depends on the
tetrahedron's six edge lengths, which are assumed to be quantized,
taking values proportional to $2j + 1$ where $j$ is a half-integer
spin.  One can compute this amplitude either by evaluating an $\SU(2)$
spin network shaped like a tetrahedron, or by doing an integral.
Approximating this integral by the stationary phase method, Ponzano
and Regge argued that when all six spins are rescaled by the same
factor $\smult$, the $\smult \to \infty$ asymptotics of the amplitude
are given by a simple function of the volume of the tetrahedron and
the Regge calculus version of its Einstein action.  Nobody has yet
succeeded in making their argument rigorous, but Roberts
\cite{Roberts,Roberts2} recently proved their asymptotic formula by a
different method.  This result lays the foundation for a careful study
of the relation between the Ponzano--Regge model and classical general
relativity in 3 dimensions.

Our concern here is whether a similar result holds for the
Barrett--Crane model of \emph{4-dimen\-sion\-al} Riemannian quantum
gravity~\cite{BC}.  In this model an amplitude is associated with each
4-simplex in a triangulation of spacetime.  This amplitude, known as a
$10j$ symbol, is a function of the areas of the 10 triangular faces of
the 4-simplex.  Each triangle area is proportional to $2j + 1$, where
$j$ is a spin labelling the triangle.  The amplitude can be
computed by evaluating an $\SU(2) \times \SU(2)$ 
spin network whose edges correspond to the triangles of the 4-simplex:
\[   \TenJ  \]
There is also an integral formula for the $10j$ symbol~\cite{Barrett}.  
The problem is to understand the asymptotics of the $10j$ symbol as all 
10 spins are rescaled by a factor $\smult$ and $\smult \to \infty$.

Barrett and Williams~\cite{BW} applied a stationary phase approximation
to the integral for the $10j$ symbol, focusing attention on stationary
phase points corresponding to nondegenerate 4-simplices with the
specified face areas.  They showed that each such 4-simplex contributes
to the $10j$ symbol in a manner that depends on its Regge action.  They
pointed out the existence of contributions from degenerate 4-simplices,
but did not analyse them. 

In Section 2 of this paper we begin by applying Barrett and Williams'
estimate of the $10j$ symbols to the case where all 10 spins are equal.
We find that the contribution of their stationary phase points
to the $10j$ symbols is of order $\smult^{-9/2}$.  However, our numerical
calculations show that the $10j$ symbols are much larger, of order
$\smult^{-2}$.  This means we must look elsewhere to explain the
asymptotics of the $10j$ symbol.  

In Section 3 we analyse the contribution of `degenerate 4-simplices'
to the integral for the $10j$ symbols.  They do not correspond to
stationary phase points in the integral for the $10j$ symbols;
instead, the integrand has a strong {\it maximum} at these points.  We
argue that the contribution of a small neighborhood of these points is
asymptotically proportional to $\smult^{-2}$.  We give a formula
expressing the constant of proportionality as an integral over the
space of degenerate 4-simplices.  We also reduce this to an explicit
integral in 5 variables.

In Section 4, we numerically compare these results to the $10j$ symbols 
as calculated using the algorithm developed by Christensen and Egan
\cite{Algorithm}.  Our formula for the contribution of degenerate
4-simplices closely matches the actual asymptotics of the $10j$
symbols.  Thus, even though our argument that these asymptotics are
dominated by degenerate 4-simplices is not rigorous, we feel confident
that the resulting formula is correct.  
 
In Section 5 we discuss a new sort of spin network, associated to the
representation theory of the Euclidean group, which arises naturally
in our analysis of the contribution of degenerate 4-simplices.
Generalizing our results on the Riemannian $10j$ symbols, we
conjecture formulas for the asymptotics of a large class of Riemannian
spin networks in terms of these new `degenerate spin networks'.  We
verify this conjecture in a number of simple cases.

In Section 6 we formulate a similar conjecture for Lorentzian spin
networks.  Taken with the previous one this conjecture implies that
the $\smult \to \infty$ asymptotics of a Lorentzian spin network in
this class are the same, up to a constant, as those of the
corresponding Riemannian spin network.  For example, as $\smult \to
\infty$, the Lorentzian $10j$ symbol should be asymptotic to $1/16$
times the corresponding Riemannian $10j$ symbol!  We conclude by
presenting some numerical evidence that this is the case.  

\section{Stationary Phase Points}
\label{S:Stationary}

The $10j$ symbols can be defined using a Riemannian spin network ---
also known as a `balanced' spin network~\cite{BC} --- whose underlying
graph is the complete graph on five vertices.  The ten edges of the
graph are labelled with half-integer spins $j_{kl} =
0,\frac{1}{2},1,\dots$, where $k$ and $l$ refer to the vertices
connected by each edge.  In this approach, an edge labelled by the
spin $j$ corresponds to the representation $j \otimes j$ of $\Spin(4)
= \SU(2) \times \SU(2)$, and the $10j$ symbols are computed using the
representation theory of this group.  

However, to analyse the
asymptotics of the $10j$ symbol, it is easier to use the integral
formula due to Barrett~\cite{Barrett}.  This is:
\begin{equation}
\label{E:riemannianIntegral}
\begin{split}
  \TenJ 
  = 
  (-1)^{\sum_{k<l}2j_{kl}}
  \int_{(S^3)^5} \prod_{k<l} \riemannianKernel{2j_{kl}+1}{\phi_{kl}}
  ~\frac{dh_1}{2\pi^2} \cdots \frac{dh_5}{2\pi^2},
\end{split}
\end{equation}
where $S^3$ is the unit 3-sphere in $\R^4$ equipped with its usual
Lebesgue measure, $\phi_{kl}$ is the angle between the unit vectors
$h_k$ and $h_l$, and the kernel $K^R$ is given by:
\begin{equation}
\label{E:riemannianKernel}
\riemannianKernel{a}{\phi} := \frac{\sin a\phi}{\sin\phi} .
\end{equation}
The normalizing factors in the integral come from the fact that
the volume of the unit 3-sphere is $2\pi^2$.

The $10j$ symbol gives the amplitude for a 4-simplex with specified
triangle areas.   Each vertex of the above graph corresponds to a
tetrahedron in this 4-simplex, and each edge of the graph corresponds to
the unique triangle shared by two of these tetrahedra.  In this picture,
the spin $j_{kl}$ determines the area of the triangle shared by the
$k$th and $l$th tetrahedra.  The precise formula for this area is
% A driveway is shared by the first and second houses on the street.
somewhat controversial \cite{Ashtekar,APB}, but given the integral 
formula for the $10j$ symbols, we find it convenient to assume the
area is proportional to $2j_{kl} + 1$.  Ignoring the constant factor, 
we thus define triangle areas by:
\[         a_{kl} = 2j_{kl} + 1 . \]

In what follows, we study the behaviour of the $10j$ symbol as all these
triangle areas $a_{kl}$ are multiplied by a large integer $\smult$.   
This is not the same as multiplying the spins $j_{kl}$ by $\smult$.
However, note that as $j_{kl}$ ranges over all spins, $a_{kl}$ ranges 
over all positive integers.   This means that if we multiply the areas
$a_{kl}$ by any positive integer $\smult$, we can find new spins
$J_{kl}$ corresponding to the new areas by solving 
$\smult a_{jk} = 2J_{kl} + 1$.    

It is shown in~\cite{Positivity} that the integral in
~\eqref{E:riemannianIntegral} is nonnegative.  Accordingly, we will
concentrate on analysing the asymptotics of the absolute value of the
$10j$ symbol, given by the integral alone:
\begin{equation}
\label{E:absoluteValue}
  \left| \; \Ten \;\;\right|
  := 
   \int_{(S^3)^5} 
  \prod_{k<l} \riemannianKernel{\smult a_{kl}}{\phi_{kl}}
  ~\frac{dh_1}{2\pi^2} \cdots \frac{dh_5}{2\pi^2} ,
% I guess there's a colon above because we're using J_{kl} not j_{kl}?
\end{equation}
where for simplicity we have left out the spins labelling the
spin network edges, which are now $J_{kl}$.

Barrett and Williams~\cite{BW} express the numerator in the kernel
$K^R$ as a difference of exponentials, which allows
them to rewrite the integrand in equation~\eqref{E:absoluteValue}
as a sum of $2^{10}$ terms, each consisting of an exponential times the function:
\begin{equation}
\label{E:statPhaseF}
\f(h_1,\dots,h_5) = \prod_{k<l} \frac{1}{\sin \phi_{kl}}.
\end{equation}
This function is unbounded as any of the $\phi_{kl}$ tend to zero or
$\pi$, but if the regions of the domain where that occurs are set
aside for a separate analysis, the integral over the remaining region
becomes amenable to a stationary phase approximation~\cite{StatPhase}.
The relevant phase is simply:
\begin{equation} 
\label{E:statPhaseG} 
S(h_1,\dots,h_5) = \sum_{k<l} \smult a_{kl} \phi_{kl}  .
\end{equation}   
The phases of the $2^{10}$ exponentials arising from the product of the kernels include
variants of $S$ with all possible signs for the ten
terms.  However, Barrett and Williams use the invariance of the
integrand under the transformations $h_k \to -h_k$ to break the integral
into $2^5$ identical parts, according to whether $h_k=\pm n_k$, where
$n_k$ are the outward normals of a 4-simplex whose five tetrahedra lie in
the hyperplanes normal to the $h_k$.
They then focus on the region where all the $h_k=n_k$, and 
show that of the $2^{10}$ phases, only $S$ and its opposite
have stationary points in this region, and they occur when the $\phi_{kl}$
are the angles between the outward normals to the tetrahedra of a
4-simplex whose ten faces have areas given by $\smult a_{kl}$.  In
this case the $h_k$ can be interpreted as the outward normals to the
five tetrahedra, the $\phi_{kl}$ are the angles between these normals,
and $\smult a_{kl}$ is the area of the face shared by the tetrahedra
numbered $k$ and $l$.  Interpreted this way, $S$ is
precisely the Regge action for the 4-simplex.

This means that the relevant stationary phase points are those of a much simpler
integral:
\begin{equation}
\label{E:statPhaseIntegral}
\begin{split}
  \frac{1}{2^5} \int \f(h_1,\dots,h_5)
  (e^{i S(h_1,\dots,h_5)} + e^{-i S(h_1,\dots,h_5)})
  ~\frac{dh_1}{2\pi^2} \cdots \frac{dh_5}{2\pi^2} .
\end{split}
\end{equation}
To carry out a stationary phase approximation of this integral we must
note that the stationary phase `point' determined by the geometry of a
4-simplex is not actually a single point in the 15-dimensional
manifold $(S^3)^5$, but rather a whole 6-dimensional submanifold,
since the geometry of the 4-simplex is invariant under the 6-parameter
rotation group $\SO(4)$.  However, the same invariance can be used to
remove this complication.  Since the functions $f$ and $S$ are
invariant under the action of $\SO(4)$, they pass to the 
quotient space $(S^3)^5/\SO(4)$, and we obtain:
\begin{equation}
\label{E:statPhaseIntegralReduced}
\begin{split}
  \frac{1}{2(2\pi)^{10}} \int_{(S^3)^5/\SO(4)} 
  \f(x) (e^{i S(x)} + e^{-i S(x)}) ~d\mu(x) ,
\end{split}
\end{equation}
where $d\mu$ is the result of pushing forward Lebesgue measure 
on $(S^3)^5$ to this quotient space.  A factor of $1/2$ here accounts
for the complication that simultaneously negating all the $h_k$ has the same action
as $-I$, an element of $\SO(4)$, and so the discrete symmetries used to
obtain~\eqref{E:statPhaseIntegral} already entailed taking the quotient by
a 2-element subgroup of $\SO(4)$.

Now, the stationary phase approximation~\cite{StatPhase} of an
$n$-dimensional integral
\[
\int \f(x) e^{i S(x)} \, dx  
\]
is a sum over stationary points $x_i$ of the function $S$:
\[
(2\pi)^{n/2} \sum_i  \frac{f(x_i)}{|\det{H(x_i)}|^{1/2}} 
  \exp\left[i S(x_i) + \frac{i\pi}{4} \sigma(H(x_i)) \right] ,
\]
where $H(x_i)$ is the matrix of second partial derivatives of $S$ at
the point $x_i$, and $\sigma(H(x_i))$ is the signature of this matrix,
i.e., the number of positive eigenvalues minus the number of negative
eigenvalues.  This approximation assumes there are finitely many
stationary points, all with $\det{H(x_i)} \ne 0$.  Applying this to
the case at hand, and neglecting points where some of the angles
$\phi_{kl}$ are $0$ or $\pi$, we obtain Barrett and Williams'
stationary phase approximation of the $10j$ symbols:
\begin{equation}
\label{E:statPhaseGeneralResult}
  \left| \; \Ten \; \;\right|_{\rm stat} :=
 \frac{1}{(2\pi)^{11/2}} \sum_i  \frac{f(x_i)}{|\det{H(x_i)}|^{1/2}} 
  \cos\left[ S(x_i) + \frac{\pi}{4} \sigma(H(x_i)) \right],
\end{equation}
where the sum is over the 4-simplices $x_{i}$ with the specified
face areas.

We can easily say something about the $\smult \to \infty$ behavior of 
this quantity without actually evaluating it.  Each stationary point
$x_i$ is independent of $\smult$, so $f(x_i)$ will be constant as
$\smult \to \infty$, while $S(x_i)$ will grow linearly, as will its
matrix of second derivatives, $H(x_i)$.  This is a $9 \times 9$ matrix,
since $(S^3)^5/\SO(4)$ is 9-dimensional, so its determinant will grow 
as $\smult^9$.  It follows that:
\begin{equation} 
\label{E:statPhaseGeneralEstimate}  
\left| \; \Ten \; \; \right|_{\rm stat} = O(\smult^{-9/2}) .
\end{equation} 
However, cancellation between different stationary points could in
principle make this a misleading overestimate.  Thus it seems worthwhile
to explicitly evaluate the sum in equation~\eqref{E:statPhaseGeneralResult}, 
at least in a simple special case.

In what follows we evaluate this sum for the case of a $10j$ symbol with
all edges labelled by the same spins.  In other words, we calculate
the stationary phase approximation of the amplitude for a regular
4-simplex.  

The first step, which would be useful more generally, is to find
explicit coordinates on the quotient space $(S^3)^5/\SO(4)$ and
describe the measure $d\mu$ in these terms.  To do this, we exploit
the fact that any configuration of the five unit vectors $h_k$ can be
rotated into one that belongs to a 9-dimensional subspace of the
original domain.  Specifically, if we express the vectors $h_k$ in
polar coordinates
\[
h_k = (\cos\psi_k, \, \sin\psi_k \cos\theta_k, \,
       \sin\psi_k \sin\theta_k \cos\phi_k, \, \sin\psi_k \sin\theta_k \sin\phi_k) ,
\]
where $0\leq\psi_k,\theta_k\leq\pi$ and $0\leq\phi_k\leq 2\pi$, any set
of $h_k$ can be rotated in such a way that the following restrictions
are met:
\begin{subequations}
\label{E:gaugeFix}
\begin{align}
h_1 &= (1,~0,~0,~0) &\psi_1=\theta_1=\phi_1=0 \\
h_2 &= (\cos\psi_2,~\sin\psi_2,~0,~0) &\theta_2=\phi_2=0 \\
h_3 &= (\cos\psi_3,~\sin\psi_3 \cos\theta_3,~\sin\psi_3 \sin\theta_3,~0)  
        &\phi_3=0 .
\end{align}
\end{subequations}
By means of this `gauge-fixing' we can use the remaining 9 variables
$\psi_2,$ $\psi_3,$ $\psi_4,$ $\psi_5,$ $\theta_3,$ $\theta_4,$ $\theta_5,$ 
$\phi_4,$ $\phi_5$ as coordinates on the quotient space.  
To describe the measure $d\mu$ in these coordinates recall that in
polar coordinates, Lebesgue measure on the unit 3-sphere is given by
\begin{subequations}
\label{E:measures}
\begin{equation}
dh_k={\sin}^2\psi_k \, \sin\theta_k \, d\psi_k \, d\theta_k \, d\phi_k .
\end{equation}
Since $h_1$ is completely fixed we omit $dh_1$ from the formula
for $d\mu$, only inserting a factor $2\pi^2$ due to the volume
of the 3-sphere.  Since $h_2$ and $h_3$ are partially fixed we replace
$dh_2$ and $dh_3$ by
\begin{align}
\tilde{dh}_2&= 4 \pi \, {\sin^2}\psi_2 \, d\psi_2 \\
\tilde{dh}_3&= 2 \pi \, {\sin^2}\psi_3 \, \sin\theta_3 \, d\psi_3 \, d\theta_3 .
\end{align}
\end{subequations}
We thus obtain 
\begin{equation}
\label{E:quotientMeasure}
d\mu = 2\pi^2 \, \tilde{dh}_2 \, \tilde{dh}_3 \,  dh_4 \,  dh_5 .
\end{equation}

Next, we determine the stationary points of the function $S$ in the
case where all ten spins are equal, neglecting points where some of
the angles $\phi_{kl}$ are $0$ or $\pi$ --- that is, where some of the
vectors $h_k$ are parallel or anti-parallel.  Recall that these
stationary points correspond to nondegenerate 4-simplices having all
10 face areas equal.  One obvious candidate is the regular 4-simplex.
However, we must rule out the possibility that there are other,
\emph{non-regular} 4-simplices for which all the faces have identical
areas.  Since we are excluding degenerate 4-simplices from the current
analysis, we can appeal to a theorem of Bang~\cite{Bang} which states
that if all the faces of a non-degenerate tetrahedron have the same
area, they are all congruent.  It follows that if all ten triangles in
a non-degenerate 4-simplex have the same area, they too are all
congruent.  Now, each of the ten edges of a 4-simplex is shared by
three triangles, so we can treat each edge as a triple of congruent
line segments that happen to be superimposed, giving a total of 30 in
all.  If all the triangles were congruent isoceles triangles, each
with one side of length $L$, then ten of these 30 line segments would
be of length $L$.  However, there is no way to partition ten line
segments into congruent triples.  The same argument rules out scalene
triangles. So the faces must be equilateral, and the 4-simplex must be
regular.

We must therefore find all sets of unit vectors $h_k$ which satisfy
the gauge-fixing conditions in equation~\eqref{E:gaugeFix} 
and form the outward normals of a regular 4-simplex.   One choice is: 
\[
\begin{split}
n_1 &= (1,0,0,0) \\
n_2 &= \textstyle{(-\frac{1}{4},\frac{\sqrt{15}}{4},0,0)} \\
n_3 &= \textstyle{(-\frac{1}{4}, -\frac{\sqrt{5/3}}{4},\sqrt{5/6},0)} \\
n_4 &= \textstyle{(-\frac{1}{4}, -\frac{\sqrt{5/3}}{4}, 
         -\frac{\sqrt{5/6}}{2}, -\frac{\sqrt{5/2}}{2})} \\
n_5 &= \textstyle{(-\frac{1}{4}, -\frac{\sqrt{5/3}}{4},
 -\frac{\sqrt{5/6}}{2}, \frac{\sqrt{5/2}}{2})} .
\end{split}
\]
These vectors have mutual dot products of $-\frac{1}{4}$.
They represent a single point in our 9-dimensional domain:
\begin{subequations}
\label{E:statPhasePoint}
\begin{align}
\psi_2=\psi_3=\psi_4=\psi_5=\textstyle{\cos^{-1}(-\frac{1}{4})} \\
\theta_3=\theta_4=\theta_5= \textstyle{\cos^{-1}(-\frac{1}{3})} \\
\phi_4=\textstyle{-\cos^{-1}(-\frac{1}{2})=-\frac{2\pi}{3}} \\
\phi_5=\textstyle{\cos^{-1}(-\frac{1}{2})=\frac{2\pi}{3}} .
\end{align}
\end{subequations}
The only other choice comes from interchanging $\phi_4$ and
$\phi_5$.  This yields another regular 4-simplex.

Now we take the phase ~\eqref{E:statPhaseG} and specialise to the case where
all the spins $j_{kl}$ are equal, say to $j$.  Setting $a = 2j + 1$, 
and working in our chosen coordinates, we obtain
\begin{equation}
S(\psi_2,\psi_3,\psi_4,\psi_5,\theta_3,\theta_4,\theta_5,\phi_4,\phi_5) = 
\smult a \sum_{k<l} \cos^{-1}(h_k\cdot h_l)
\end{equation}
The partial derivatives of $S$ are all zero at the point described by
~\eqref{E:statPhasePoint}.  Symbolic computer calculations show that
at this point the matrix of second derivatives of $S$
has 5 positive eigenvalues and 4 negative ones, and a determinant of:
\[ \textstyle{\sqrt{\frac{5}{3}} \cdot \frac{5}{2} \, (\frac{2}{3})^{11}}  
\, (\smult a)^9.  \]
Applying this result to~\eqref{E:statPhaseGeneralResult} at the point
described by ~\eqref{E:statPhasePoint}, working in our chosen
coordinates, redefining the function $\f$ to include
the measure~\eqref{E:quotientMeasure} as well as the
kernel denominators~\eqref{E:statPhaseF},
and multiplying by a factor of two to account for the second point
where $\phi_4$ and $\phi_5$ are interchanged, we obtain:
\begin{equation}
\label{E:statPhaseResult} 
\left| \; \Ten \; \; \right|_{\rm stat} = 
\frac{24(\frac{3}{5})^{\frac{3}{4}}}{5\pi^{\frac{3}{2}}} (\smult a)^{-9/2} 
\cos \left[10 \cos^{-1}(-\frac{1}{4})\, \smult a +
\frac{\pi}{4}\right]
\end{equation}
The above expression consists of an oscillating term times a function
proportional to $\smult^{-9/2}$.  In short, the estimate
\[
\left| \; \Ten \; \; \right|_{\rm stat} = O(\smult^{-9/2}) 
\]
is sharp, at least for the case of 10 equal spins.

However, our numerical calculations of the $10j$ symbols exhibit very
different behaviour.  Rather than being of order
$\smult^{-9/2}$, they appear to be much larger, of order
$\smult^{-2}$.  Also, they exhibit no discernable oscillations.  In
the next section we argue that these results are explained by the
contribution of `degenerate 4-simplices'.

\section{Degenerate Points}
\label{S:Degenerate}

The absolute value of the kernel  
\begin{equation}
\label{E:riemannianKernel2}
 \riemannianKernel{a}{\phi} = \frac{\sin a\phi}{\sin\phi} 
\end{equation}
has maxima when the angle $\phi$ equals $0$ or $\pi$, and these maxima
become ever more sharply peaked as $a \to \infty$.  This means that as
$\smult \to \infty$, the integrand in equation~\eqref{E:absoluteValue}
becomes very large at `degenerate points', where some of the vectors
$h_k$ are either parallel or anti-parallel.  We have made a detailed
study of the `fully degenerate' points, where {\it all} the vectors
$h_k$ are either parallel or anti-parallel.  It is plausible to expect
the contribution from a neighborhood of these points to dominate the
integral for the $10j$ symbol, at least asymptotically as $\smult \to
\infty$, since in this region we are integrating a product of kernels,
all of which are near their greatest possible absolute value.  In the
next section we shall present numerical evidence that this is in fact
the case.

The value of the kernel is positive at $\phi = 0$, but at $\phi = \pi$
its sign is positive when $2j$ is even and negative otherwise, where
$a = 2j+1$.  At first glance, this appears to allow the possibility
that parallel and anti-parallel degenerate points might cancel each
other for certain values of the spins.   In fact, cancellation
can occur only when the $10j$ symbol vanishes.  The
integrand in ~\eqref{E:riemannianIntegral} is the product of ten
kernels, so it will be positive when all the $h_k$ are parallel.  If
$h_1$ is replaced by its opposite, the four $\phi_{1l}$ will change
from zero to $\pi$, leading to an overall sign change of
$(-1)^{2(j_{12}+j_{13}+j_{14}+j_{15})}$.  However, for the $10j$
symbol to be non-zero, the spins at each vertex must sum to an
integer~\cite{BC}.  Therefore, if the $10j$ symbol is non-zero, 
the sign of the integrand will remain
positive.  The same argument applies if any subset of the $h_k$ are
reversed.

In the integral in equation~\eqref{E:absoluteValue}, fully
degenerate points occur in 3-dimensional submanifolds of the domain,
but we can apply the same gauge-fixing principles as we used to analyse
the stationary points, in this case simply fixing $h_1=(1,~0,~0,~0)$. 
There are then 16 discrete fully degenerate points in the restricted
domain: those where $h_k=\pm h_1$ for $k=2,\dots,5$.  Restricting
the integral to the vicinity of these points we obtain
\begin{equation}
\label{E:degenerateContribution}
\left| \; \Ten \; \; \right|_{\rm deg} := 
  16 \int_{U} \prod_{k<l} \riemannianKernel{\smult a_{kl}}{\phi_{kl}}
  ~\frac{dh_2}{2\pi^2} \cdots \frac{dh_5}{2\pi^2} ,
\end{equation}
where $U$ is a small open ball around the point 
$(h_1,h_1,h_1,h_1) \in (S^3)^4$.

We approximate this quantity by noting that when $\phi$ is small, the
kernel in~\eqref{E:riemannianKernel2} is close to:
\begin{equation}
\label{E:euclideanKernel}
\euclideanKernel{a}{\phi} := \frac{\sin a\phi}{\phi}
\end{equation}
We call this quantity the `degenerate kernel'.  As we shall see in
Section~\ref{S:Euclidean}, this is the analog of the original kernel
in a spin-network formalism where the space of constant curvature,
$S^3$, is replaced by three-dimensional Euclidean space, and the group
$\SO(4)$, the isometry group of $S^3$, is replaced by the Euclidean
group of isometries of $\R^3$.  

Similarly, the integral over a small subset of $(S^3)^4$ can be
approximated by an integral over a subset of $(\R^3)^4$, and the angle
$\phi_{kl}$ between unit vectors $h_k$ and $h_l$ in $S^3$ replaced by
the Euclidean distance $r_{kl}=|x_k-x_l|$ between vectors $x_k$ and
$x_l$ in $\R^3$.  In terms of these new Euclidean variables, the
restriction $h_1=(1,~0,~0,~0)$ is replaced by $x_1=(0,~0,~0)$.

Thus, when $\smult$ is large, we have:
\begin{equation}
\label{E:euclideanIntegral}
\left| \; \Ten \; \; \right|_{\rm deg} \approx
  16\int_{U}
  \prod_{k<l} \euclideanKernel{\smult a_{kl}}{|x_k-x_l|} 
~\frac{dx_2}{2\pi^2} \cdots \frac{dx_5}{2\pi^2} ,
\end{equation}
where now $U$ is a small open ball around the origin of $(\R^3)^4$.

The approximation~\eqref{E:euclideanIntegral} exhibits very
simple scaling behaviour.  First, note that the degenerate
kernel~\eqref{E:euclideanKernel} obeys the identity:
\begin{equation}
\label{E:euclideanKernelScaling}
\begin{split}
\euclideanKernel{\smult a}{r}  & =
\frac{\sin \smult a r}{r} \\
  & = \smult \euclideanKernel{a}{\smult r}
\end{split}
\end{equation}
This allows the scaling of~\eqref{E:euclideanIntegral} to be deduced
from a linear change of variables, $y_k=\smult x_k$:
\begin{equation}
\label{E:euclideanIntegralScaling}
\begin{split}
  16\int_{U}
  \prod_{k<l} \euclideanKernel{\smult a_{kl}}{|x_k-x_l|} 
~\frac{dx_2}{2\pi^2} \cdots \frac{dx_5}{2\pi^2}  \\
 =  16 \smult^{-2}\int_{\smult U}
  \prod_{k<l} \euclideanKernel{a_{kl}}{|y_k-y_l|} 
  ~\frac{dy_2}{2\pi^2} \cdots \frac{dy_5}{2\pi^2} ,
\end{split}
\end{equation}
where $\smult U$ is the result of rescaling the neighborhood $U$
by a factor of $\smult$.   If the integral on the right hand side of
this equation converges in the limit $\smult\to\infty$, we obtain
the asymptotic formula:
\begin{equation}
\label{E:euclideanIntegralLimit}
\left|\; \Ten\; \; \right|_{\rm deg} \sim 
  ~16\smult^{-2} \int_{(\R^3)^4}
  \prod_{k<l} \euclideanKernel{a_{kl}}{|y_k-y_l|} 
  ~\frac{dy_2}{2\pi^2} \cdots \frac{dy_5}{2\pi^2} .
\end{equation}
We call this integral the `degenerate $10j$ symbol'.  Apart from the
factor of $16\smult^{-2}$, the integral here can be interpreted as
the evaluation of a spin network with edges labelled by unitary
irreducible representations of the Euclidean group: the group of
isometries of Euclidean 3-space.  We discuss this interpretation in
more detail in Section~\ref{S:Euclidean}.

While elegant, the integral in \eqref{E:euclideanIntegralLimit} is
difficult to compute numerically, very much like the integral
for the Lorentzian $10j$ symbol.  To obtain a more convenient form
for evaluation, we make use of the Kirillov trace formula:
\[
 \frac{\sin|x|}{|x|} = \frac{1}{4\pi}\int_{S^2} \exp(i x \cdot \xi) ~d\xi ,
\]
where $x$ is a vector in $\R^3$, and the integral is over the unit
2-sphere with its standard measure.  This formula is easily confirmed
by choosing spherical coordinates such that the $z$-axis is parallel to
the vector $x$.  For our purposes we shall rewrite it as follows:
\begin{equation}
\label{E:KirillovKernel}
\euclideanKernel{a}{|x|} = 
\int_{S(a)} \exp(i x \cdot \xi) ~d\xi ,
\end{equation}
where $S(a)$ is the 2-sphere of radius $a$ embedded in $\R^3$, but
where $d\xi$ is the induced Lebesgue measure divided by $4\pi a$, to hide 
some annoying constants that would otherwise appear in this formula.  Using 
this we can rewrite~\eqref{E:euclideanIntegralLimit} as:
\begin{equation}
\label{E:calculation}
\begin{split}
\left|\; \Ten\; \; \right|_{\rm deg}  &\sim 
 \frac{16\smult^{-2}} {(2\pi^2)^4} 
    \int_{(\R^3)^4} \int_\X
    \exp(i \sum_{k<l} (y_k-y_l)\cdot \xi_{kl})
  ~d\xi_{12} \cdots d\xi_{45} ~dy_2 \cdots dy_5 \\
&= \frac{\smult^{-2}}{\pi^8} 
    \int_{(\R^3)^4} \int_\X
    \exp(i (y_2,y_3,y_4,y_5)\cdot F(\xi_{12}, \dots \xi_{45}))
    ~d\xi_{12} \cdots d\xi_{45} ~dy_2 \cdots dy_5 \\
&= \smult^{-2} 2^{12}\pi^4
   \int_{\X} 
   ~\delta^{12}(\F(\xi_{12},\dots,\xi_{45}))~d\xi_{12} \cdots d\xi_{45} ,
\end{split}
\end{equation}
where 
\[  \X = \prod_{k < l} S(a_{kl})  \]
is a Cartesian product of 2-spheres with the measures described above, 
and $\F \maps \X \to \R^{12}$ is defined by:
\[
\begin{split}
\F(\xi_{12},\dots,\xi_{45})=(
   & - \xi_{12} + \xi_{23} + \xi_{24} + \xi_{25}, \\
   & - \xi_{13} - \xi_{23} + \xi_{34} + \xi_{35}, \\
   & - \xi_{14} - \xi_{24} - \xi_{34} + \xi_{45}, \\
   & - \xi_{15} - \xi_{25} - \xi_{35} - \xi_{45}).
\end{split}
\]
If 
\[ \N=\{\xi\in \X :\; \F(\xi)=0\} \]
then the final integral in~\eqref{E:calculation} will be well-behaved
so long as at each point $\xi\in\N$ the differential of $\F$ has
maximal rank, namely 12.  When this is the case $\N$ is an
8-dimensional submanifold of $\X$, and the integral reduces to:
\begin{equation}
\label{E:reducedIntegral}
  \left| \;\Ten \;\; \right|_{\rm deg} \sim 
  ~\smult^{-2}\, 2^{12} \pi^4
  \int_{\N} |J(\xi)|^{-1}~d\xi
\end{equation}
Here $d\xi$ is the Lebesgue measure on $\N$ induced by the
Riemannian metric on $\X$, but divided by a factor of $\prod_{k < l} 4
\pi a_{kl}$, since we have divided the Lebesgue measure on each sphere
by a factor of $4 \pi a_{kl}$.  If we choose local coordinates
$(x^1,\dots,x^{20})$ on $\X$ near $\xi\in\N$ such that $x^1,\dots,x^8$
are zero on $\N$, then $|J(\xi)|$ is the Jacobian determinant of $\F$
as a function of the remaining 12 variables $x^9,\dots,x^{20}$.

One way to get solutions of $\F = 0$ is to start with 
4-simplices in $\R^3$: that is, 5-tuples of points in $\R^3$, together
with the ten triangles and five tetrahedra determined by these points.  Given
such a 4-simplex, let $\xi_{kl}$ be the vector that is normal to the 
triangle shared by the $k$th and $l$th tetrahedra, and has length equal 
to the area of this triangle.  As shown in ~\cite{Tetrahedron}, the 
four vectors $\xi_{kl}$ normal to the triangles in any one tetrahedron 
must sum to zero, with appropriate signs:
\begin{equation}
\label{E:constraint}
\begin{array}{rcl}
 - \xi_{12} + \xi_{23} + \xi_{24} + \xi_{25} &=& 0 \\
 - \xi_{13} - \xi_{23} + \xi_{34} + \xi_{35} &=& 0 \\
 - \xi_{14} - \xi_{24} - \xi_{34} + \xi_{45} &=& 0 \\
 - \xi_{15} - \xi_{25} - \xi_{35} - \xi_{45} &=& 0 \\
   \xi_{12} + \xi_{13} + \xi_{14} + \xi_{15} &=& 0.
\end{array}
\end{equation}
Each vector appears twice in these formulas, with opposite signs,
since the outwards-pointing normal to one tetrahedron is the
inwards-pointing normal to another.  The first four equations say
that $\F = 0$; the last is an algebraic consequence of the rest.
If $|\xi_{kl}| = a_{kl}$, the vectors $\xi_{kl}$ thus determine 
a point in $\N$.

This suggests that we think of points of $\N$ as `degenerate 4-simplices'.
However, not every point of $\N$ comes from a 5-tuple of points in
$\R^3$ this way.  To see this, note that two 5-tuples in
$\R^3$ determine the same point in $\N$ if they are translates of each 
other.  The space of 5-tuples modulo translation has dimension
$3 \times 5 - 3 = 12$, but the space of solutions of equation
~\eqref{E:constraint} has dimension $30 - 4 \times 3 = 18$.
Thus there are simply not enough 5-tuples of points in $\R^3$
to account for all solutions of equation ~\eqref{E:constraint}.
We shall still call points of $\N$ `degenerate 4-simplices', because
it is a useful heuristic.  However, it is important to take this
phrase with a grain of salt.

Of course, for the space $\N$ to be nonempty, there must \emph{exist}
a solution of $\F = 0$.  This imposes certain restrictions on the
numbers $a_{kl}$.  For example, there will be a solution
of $\xi_{12} + \xi_{13} + \xi_{14} + \xi_{15} = 0$ if and only if 
these `tetrahedron inequalities' hold:
\[
\begin{array}{ccl}
a_{12} &\le& a_{13} + a_{14} + a_{15} \\
a_{13} &\le& a_{12} + a_{14} + a_{15} \\
a_{14} &\le& a_{12} + a_{13} + a_{15} \\
a_{15} &\le& a_{12} + a_{13} + a_{14}. \\
\end{array}
\]
In general $\N$ will be empty if the four numbers $a_{kl}$
corresponding to the faces of any one tetrahedron violate the
tetrahedron inequalities.  In this case the degenerate $10j$
symbols vanish.  Similarly, the Riemannian $10j$ symbols vanish if 
the spins $j_{kl}$ violate the tetrahedron inequalities~\cite{BC}, 
and for the same sort of geometrical reason~\cite{Tetrahedron}.

For computational purposes it is helpful to rewrite the integral for
the degenerate $10j$ symbols in yet another way.  To do this, first we
note that we can exploit $\SO(3)$ invariance to
convert~\eqref{E:reducedIntegral} to an integral over the
5-dimensional manifold $\N/\SO(3)$, by rotating the $\xi_{kl}$ in $\X$
so that they lie in the 17-dimensional subspace where $\xi_{23}$ is
fixed at $(0,~0,~a_{23})$, and $\xi_{34}$ lies in the $x>0$ half of
the $xz$-plane.  This corresponds to integrating over all possible
values for $\xi_{23}$, inserting a factor equal to the volume of
$S(a_{23})$ with our chosen measure, which is $a_{23}$, and also
integrating over all possible azimuthal coordinates for $\xi_{34}$ and
inserting a factor of $2\pi$.

In what follows, we will choose coordinates so that the $12 \times 12$
matrix $J(\xi)$ whose determinant we require is block
diagonal, with two $3 \times 3$ blocks and one $6 \times 6$ block.

It turns out to be
convenient to parameterise $\xi_{34}$, not by its angle from the
$z$-axis, but by the length
\[
s_1:=|\xi_{23}-\xi_{34}| .
\]
Since $\xi_{13}+\xi_{23} = \xi_{34}+\xi_{35}$, these four vectors can
be positioned to form the sides of a (possibly non-planar) quadrilateral.
Then $s_1$ is the length of one of the diagonals.
With the vectors $\xi_{23}$ and $\xi_{34}$ fixed and the lengths of
$\xi_{13}$ and $\xi_{35}$ specified, the only
remaining freedom this quadrilateral has, if it is to remain closed, is
the `hinge angle', $\alpha_1$, between the two triangles that meet
along the diagonal.  The vectors $\xi_{13}$ and $\xi_{35}$ have 4 degrees
of freedom in all, and specifying $\alpha_1$ removes one of them,
leaving 3 which break the quadrilateral.  With our chosen measure on $\N$, the
product of the measure for the coordinates we are integrating over,
and the Jacobian determinant for the 3 that break the quadrilateral, is:
\[
\frac{1}{(4\pi)^3 a_{23}} .
\]
We can treat a second 4-tuple of vectors more or less identically.  If the
vector $\xi_{12}$ has an azimuthal angle of $\phi$, and we define
\[
s_2:=|\xi_{23}-\xi_{12}| ,
\]
then the quadrilateral formed by $\xi_{23}, \xi_{12}, \xi_{25}$ and
$\xi_{24}$ can be assigned a `hinge angle' of $\alpha_2$.  This
specifies all the degrees of freedom that allow this quadrilateral 
to remain closed.  Once again, the
product of the measure for the coordinates we are integrating over,
and the Jacobian determinant for the 3 that break the quadrilateral, is:
\[
\frac{1}{(4\pi)^3 a_{23}} .
\]
So far, we have specified 5 degrees of freedom for $\N/\SO(3)$: $s_1$, $s_2$,
$\alpha_1$, $\alpha_2$ and $\phi$.  No further continuous degrees of
freedom remain.  The three vectors we have yet to specify must form
triangles that complete two quadrilaterals with vectors that have
already been parameterised, and because these two triangles have a
vector in common, there is no `hinge' freedom left.

Specifically, the three vectors $\xi_{14}, \xi_{15}$ and $\xi_{45}$ must
form a tetrahedron by fitting over a triangular base that has, as two of
its sides, the vectors:
\begin{align*}
v &:= \hspace{0.7em} \xi_{35} + \xi_{25} \\
w &:=              - \xi_{24} - \xi_{34} .
\end{align*}
Given their fixed lengths, this determines $\xi_{14}, \xi_{15}$ and
$\xi_{45}$ completely, apart from the freedom to locate the apex of
the tetrahedron on either side of the plane spanned by $v$ and $w$.
This freedom can be accounted for with a factor of 2 in the integral.
The final contribution to the Jacobian comes from a $6 \times 6$ block
involving all the coordinates of $\xi_{14}, \xi_{15}$ and $\xi_{45}$,
and with our chosen measure this is:
\[
\frac{1}{(4 \pi)^3~6 \V(a_{14},a_{15},a_{45},|w|,|v-w|,|v|)} ,
\]
where $\V$ is the volume of the tetrahedron as a function of its edge 
lengths.

Combining these results, we can rewrite \eqref{E:reducedIntegral} as:
\begin{equation}
\label{E:coefficientIntegral}
  \left| \;\Ten \;\; \right|_{\rm deg} \sim 
  \frac{\smult^{-2}}{96\pi^{4} a_{23}}
  \int_{s_1} \int_{s_2} \int_{\alpha_1} \int_{\alpha_2} \int_\phi
  \frac{ds_1 \, ds_2 \, d\alpha_1 \, d\alpha_2 \, d\phi}
       {\V(a_{14},a_{15},a_{45},|w|,|v-w|,|v|)} .
\end{equation}
The integrals over the $s_i$ are taken over intervals
determined by the four sides of the quadrilaterals for which they are
the diagonal lengths, and the angular variables range from $0$ to
$2\pi$, with the proviso that any part of the domain where the
tetrahedron is not geometrically possible must be excluded.  In
numerical calculations, this can be dealt with by setting the integrand
to zero wherever Cayley's determinant formula for the squared volume of
the tetrahedron yields a negative value.

We note that the integrand here is unbounded, and we have not proved
that \eqref{E:coefficientIntegral} converges, but our numerical
calculations suggest that it does.
\vskip 30em

\section{Numerical Data}
\label{S:Numerical}

To test our hypothesis that the asymptotics of $10j$ symbols are
dominated by the contribution of degenerate 4-simplices, we used the
algorithm described in~\cite{Algorithm} to calculate values for
several sets of $10j$ symbols.  The figure below shows log-log plots
for the absolute values of the Riemannian $10j$ symbols as a function 
of $\smult$, where $\smult$ is the parameter by which the areas $a_{kl}$
were multiplied.  The legend shows the base spins $j_{kl}$; multiplying
$a_{kl}$ by $\smult$ was achieved by replacing the $j_{kl}$ with:
\[
J_{kl} = \smult j_{kl} + \frac{\smult - 1}{2} .
\]
%\enlargethispage{\baselineskip}
\vspace{-10pt}
{
\center
\includegraphics{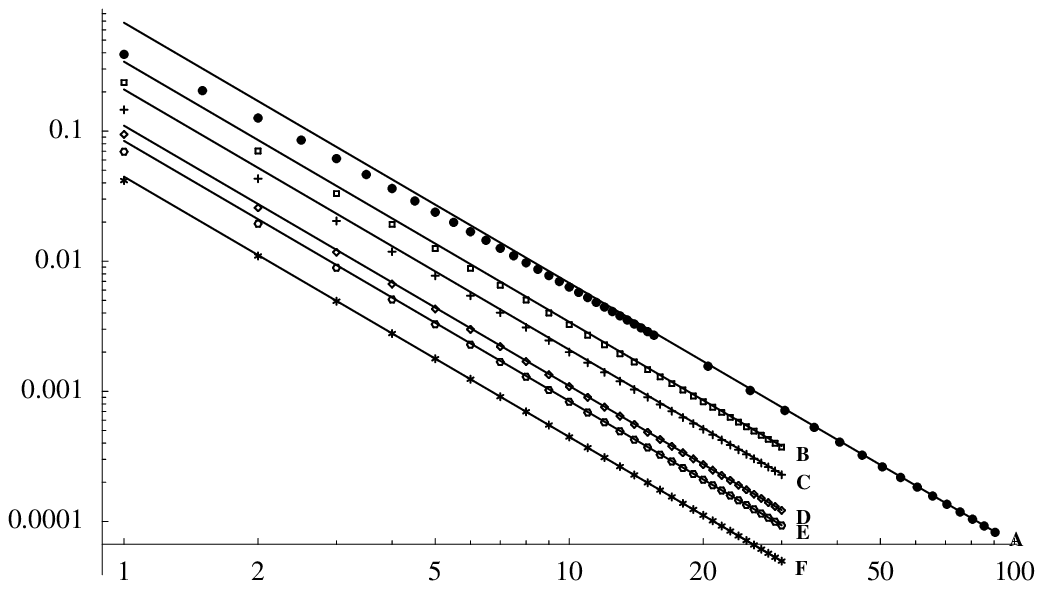}\\
\epsfig{file=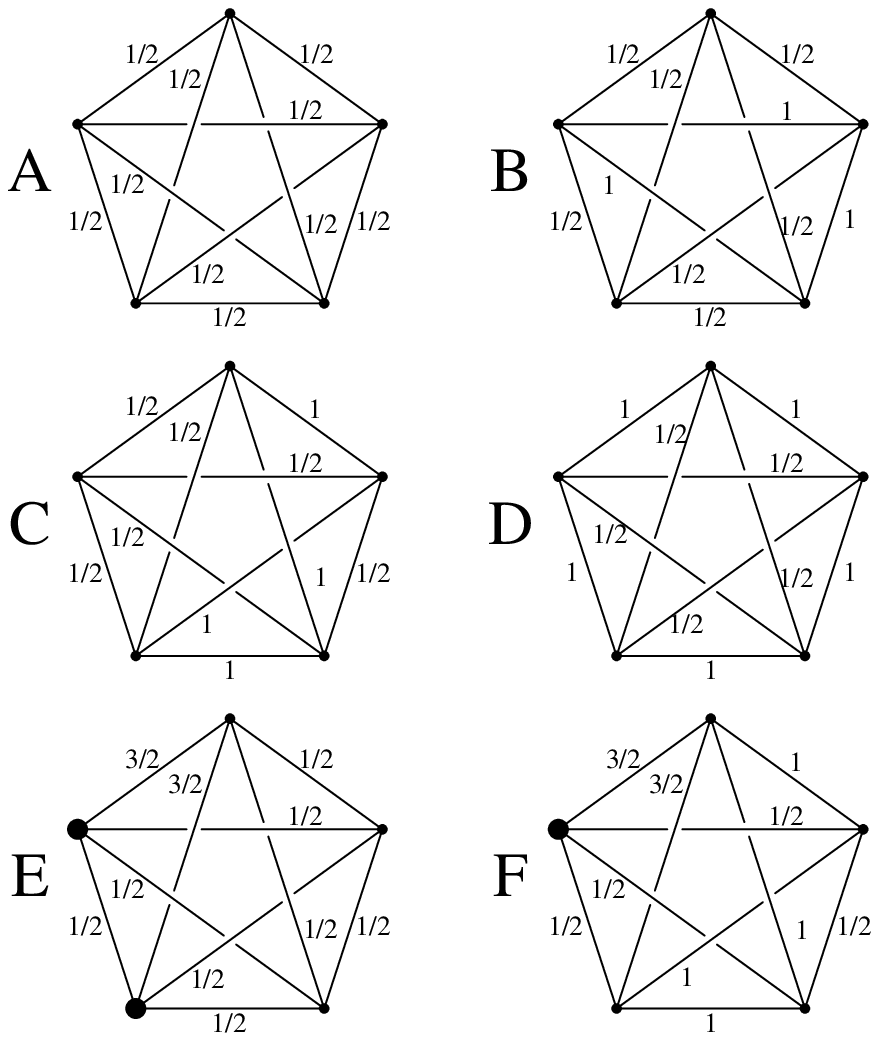,height=3.7in}\\
}
\vspace{1em}
\noindent
The lines on the plot show the 
$\smult^{-2}$ asymptotic behaviour described by equation
~\eqref{E:coefficientIntegral}; the integrals were evaluated numerically
with Lepage's VEGAS algorithm~\cite{Vegas}, and found to have values of
0.680, 0.341, 0.209, 0.110, 0.0841 and 0.0446 respectively.

In summary, our numerical data supports the following conjecture.
Let us say that the spins $j_{kl}$ are `admissible' if for each 
vertex in the $10j$ symbol, the spins labelling the four incident
edges satisfy the tetrahedron inequalities and sum to an integer.  
Then:

\begin{conj} If the ten spins $j_{kl}$ are admissible, the 
$\smult \to \infty$ asymptotics of the Riemannian $10j$ symbols 
are given by:
\[
\left|\; \Ten\; \; \right| \sim 
  ~16\smult^{-2} \int_{(\R^3)^4}
  \prod_{k<l} \euclideanKernel{2j_{kl}+1}{|y_k-y_l|} 
  ~\frac{dy_2}{2\pi^2} \cdots \frac{dy_5}{2\pi^2} .
\]
\end{conj}
\noindent 
In the next two sections we generalize this conjecture to 
a large class of Riemannian and Lorentzian spin networks, including
the Lorentzian $10j$ symbols.

The reader may have wondered why we consider asymptotics of the $10j$
symbols as \emph{areas} are rescaled, instead of \emph{spins}.  The
reason is that they are much simpler.  We also calculated values for
sets of $10j$ symbols where the spins $j_{kl}$ were multiplied by
$\smult$.  The figure below shows log-log plots for the absolute
values of the Riemannian $10j$ symbols with spins $\smult j_{kl}$,
where the $j_{kl}$ match those shown in the legend for the previous
figure:

{
\center
\includegraphics{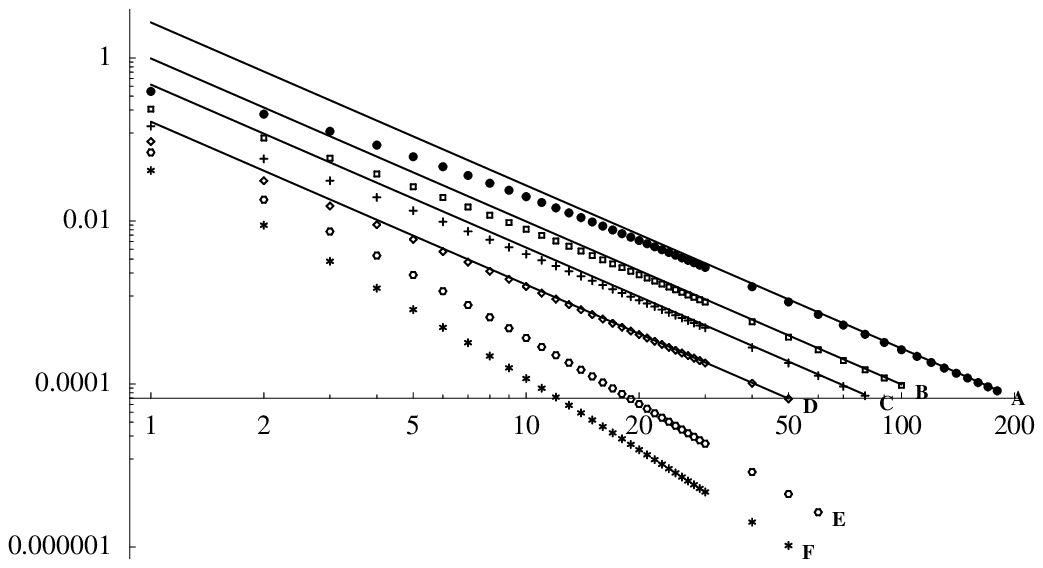}
}

\vskip 1em

Since the areas $2 \smult j_{kl} + 1$ are asymptotic to a series of
values $2 \smult j_{kl}$ that are proportional to $\lambda$, the data
might be expected to exhibit $\smult^{-2}$ scaling.  And in fact, in
cases A-D, that is exactly what is seen.  Furthermore, numerically
evaluating the integral~\eqref{E:coefficientIntegral} with $a_{kl}=2
j_{kl}$ provided the correct coefficients for lines on the plot which 
are asymptotic to the data; these coefficients were 2.73, 0.987, 0.472 and
0.165 respectively.

For cases E and F, the $10j$ symbols have one or more vertices at the
`border of admissibility': that is, vertices where three of the spins
labelling incident edges sum to equal the fourth spin, making one of
the tetrahedron inequalities a strict equality.  These vertices are
marked with heavy dots on the legend.  Setting $a_{kl}=2 j_{kl}$
in~\eqref{E:coefficientIntegral} in these cases yields a coefficient
of zero, because the domain for at least one variable of integration,
the quadrilateral diagonal length $s_1$, was reduced to a single
point.  Empirically, the data here appears to scale as $\smult^{-3}$.

While $10j$ symbols with one or two vertices on the border of
admissibility have $\smult^{-3}$ asymptotics with spin rescaling, $10j$
symbols with three or more vertices on the border of admissibility
appear to decay exponentially as their spins are multiplied by
$\smult$, as illustrated in the log-linear plot below:

{
\center
\includegraphics{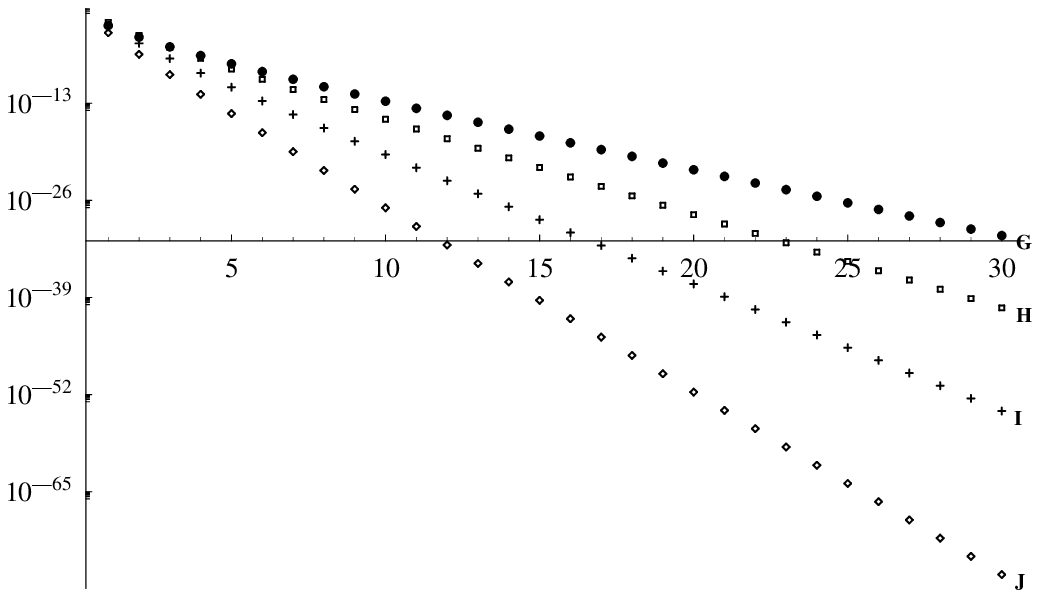}

\center \hskip 8em
\epsfig{file=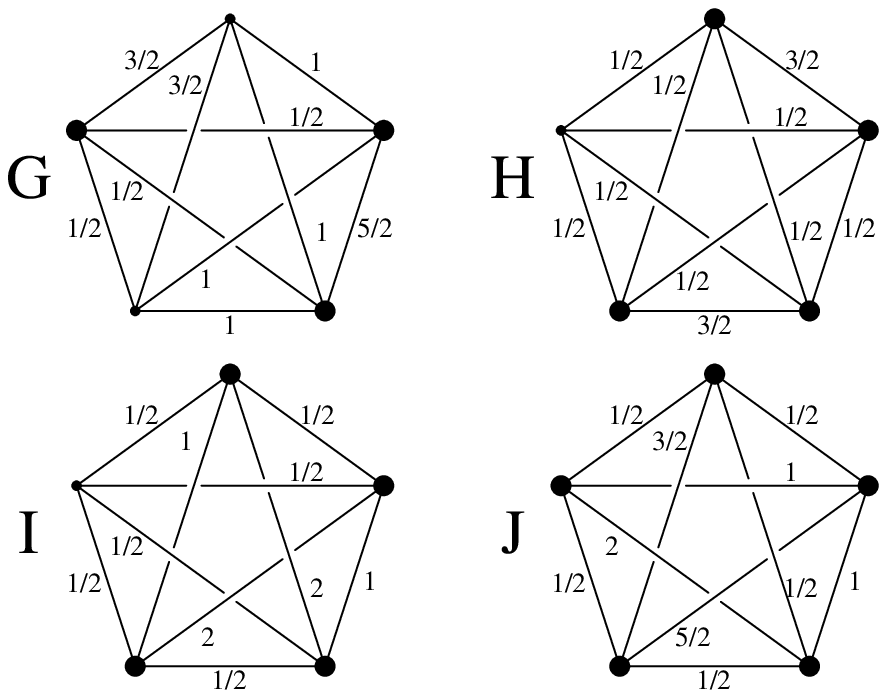,height=2.55in}
}
\vskip 1em

To prove convergence of the partition function in our newly
formulated version of the Barrett--Crane model~\cite{Riemannian}, 
it will probably be necessary to understand the $\smult^{-3}$
and exponential decay of borderline-admissible $10j$ symbols 
under spin rescaling.  However, these are more delicate phenomena
than we are prepared to tackle here.

\section{Degenerate Spin Networks}
\label{S:Euclidean}

Though the Riemannian $10j$ symbols are given by an integral
over a product of copies of $S^3$, we have seen that 
when we rescale the areas by a large constant $\smult$, this integral
is dominated by the contribution of a very small patch of this
space, which can be approximated by a product of copies of $\R^3$.
Indeed, by a change of variables we can think of the $\smult \to \infty$ 
limit as one in which the radius of $S^3$ approaches infinity,
so that it degenerates to Euclidean 3-space.

The Riemannian $10j$ symbols can also be described using the 
representation theory of $\SO(4)$, the isometry group of $S^3$. 
This suggests that the $\smult \to \infty$ asymptotics of the $10j$
symbols can be described in terms of the representation theory 
of the Euclidean group $\E(3)$, the isometry group of $\R^3$.  

To do this, we introduce certain spin networks associated to $\E(3)$
which we call `degenerate spin networks'.  These are more
closely analogous to the Lorentzian spin networks defined
in~\cite{BCL} than to the Riemannian spin networks we have been
discussing so far, because the edge labels are not restricted to discrete
values, and the group representations are infinite-dimensional.
However, all three sorts of spin network form part of a unified
theory, as outlined below:
\vskip 1em
{\vbox{   
\begin{center}   
\begin{tabular}{|c|c|c|c|}            \hline
geometry  & signature  & symmetry group & homogeneous space \\ \hline 
Riemannian& (++++)     & $\SO(4)$        & $S^3$               \\ 
degenerate& ($0$+++)     & $\E(3)$         & $\R^3$              \\ 
Lorentzian& ($-$+++)     & $\SO(3,1)$      & $H^3$               \\ \hline
\end{tabular}
\end{center}   
}}
\vskip 1em 
\noindent
The rotation group $\SO(4)$ and the Lorentz group $\SO(3,1)$ consist
of linear transformations of $\R^4$ with determinant 1 that preserve
metrics of signature (++++) and ($-$+++), respectively.  Similarly,
the Euclidean group $\E(3)$ is isomorphic to the group of linear
transformations of $\R^4$ with determinant 1 that preserve the singly
degenerate metric with signature ($0$+++).  The groups $\SO(4)$ and
$\SO(3,1)$ both `contract' to $\E(3)$, meaning that they have it
as a limiting case if one forms the isometry group of the metric ${\rm
diag}(\epsilon,1,1,1)$ and lets $\epsilon \downarrow 0$
and $\epsilon \uparrow 0$, respectively.  

This makes it plausible that the asymptotics of not only Riemannian 
but also \emph{Lorentzian} spin networks can be calculated using
degenerate spin networks.  In this section we state a precise
conjecture along these lines for a large class of Riemannian spin
networks, and present some supporting evidence.
In Section \ref{S:Lorentzian} we do the same for Lorentzian spin networks.

We begin by describing degenerate spin networks and how to evaluate
them.  The representations $j \otimes j$ labelling edges of a
Riemannian spin network are representations not just of $\Spin(4)$,
but actually of $\SO(4)$.  As emphasized by Freidel and Krasnov
\cite{FK}, these representations are precisely the eigenspaces of the
Laplacian on $S^3$, which is the homogeneous space $\SO(4)/\SO(3)$.
Similarly, the representations labelling edges of a Lorentzian spin
network are the eigenspaces of the Laplacian on hyperbolic 3-space,
$H^3 = \SO_0(3,1)/\SO(3)$ --- here we must take the connected component
of the Lorentz group to get just one sheet of the hyperboloid.
Following this pattern, the representations labelling edges of a
degenerate spin network should be the eigenspaces of the Laplacian on
$\R^3 = \E(3)/\SO(3)$.

In fact, there is one representation of this sort for each positive
real number $a$.  Any complex function on $\R^3$ with $\nabla^2
f=-a^2f$ can be written as:
\begin{equation}
\label{E:transform}
    f(x) = \int_{\xi \in S(a)} \hat{f}(\xi) ~\exp(i\xi \cdot x) ~d\xi ,
\end{equation}
where $S(a)$ is the 2-sphere of radius $a$ centered at the origin
of $\R^3$, and $d\xi$ is the induced Lebesgue measure divided by
$4\pi a$.  Defining an inner product on these solutions by 
\[  \langle f,g\rangle = 
\int_{\xi \in S(a)} \overline{\hat{f}(\xi)} ~\hat{g}(\xi) ~d\xi ,\]
we form a Hilbert space $\H_a$ consisting of all solutions $f$ with
$\langle f,f \rangle < \infty$.  This Hilbert space becomes a
representation of the Euclidean group where each group element $g$ 
acts via $(gf)(x)=f(g^{-1}x)$.  In fact, this representation is unitary 
and irreducible.  The functions $\exp(i\xi \cdot x)$ form a `basis', in
the sense that any element of $\H_a$ can be expressed as in equation
\eqref{E:transform} for a unique square-integrable function $\hat f$
on the sphere.

We define a `degenerate spin network' to be a directed graph with each
edge $e$ labelled by a representation of this form, or equivalently, a
positive number $a_e$.  An intertwiner between tensor products of
these representations can be defined at each vertex by taking the
product of functions from the representations labelling the
\emph{incoming} edges, multiplying it by the product of complex
conjugates of functions from representations labelling the \emph{outgoing}
edges, and integrating the result over $\R^3$.  Given this, the
standard way to evaluate such a spin network \cite{Foam} would be to
take a `trace': that is, integrate over a basis label $\xi_e \in
S(a_e)$ for each edge of the graph.  The result would be:
\[  
       	\int_{\!\prod\limits_{v \in V}\!\R^3} 
       \prod_{e \in E} \left[ \int_{S(a_e)} 
       \exp(i(x_{s(e)} - x_{t(e)}) \cdot \xi_e) ~d\xi_e \right]
       \; \prod_{v \in V} \frac{dx_v}{2\pi^2} .
\]
Here $E$ denotes the set of edges of the graph, $V$ denotes its set of
vertices, and the vertices $s(e)$ and $t(e)$ are the source and target
of the edge $e$.

However, just as in the Lorentzian case~\cite{BCL}, this gives a
divergent integral, because the integrand is invariant when we
simultaneously translate all the vectors $x_v \in \R^3$ by the same
amount.  More generally, if the underlying graph of our spin network
consists of several connected components, and we translate the vectors
$x_v$ where $v$ lies in any one component, the integrand does not
change.  To keep things simple, let us consider only spin networks
whose underlying graph is \emph{connected}.  In this case we can
sometimes obtain a well-defined integral by `gauge-fixing' one of the
vectors rather than integrating over it: that is, setting $x_{v_1} = x
\in \R^3$ for some vertex $v_1 \in V$.  If we let $V' = V - \{v_1\}$,
this gives the following formula for evaluating a degenerate spin
network with edges labelled by the numbers $a_e$:
\begin{equation}
\label{E:euclideanEvaluation1}  
  \euclideanIntegral{a} = \int_{\!\!\prod\limits_{v \in V'}\!\!\R^3} 
           \prod_{e \in E} \left[ \int_{S(a_e)} 
           \exp(i(x_{s(e)} - x_{t(e)}) \cdot \xi_e) ~d\xi_e \right]
           \; \prod_{v \in V'} \frac{dx_v}{2\pi^2} .
\end{equation}
As in the Lorentzian case~\cite{BBL}, one can show that if this
integral converges, the result does not depend on our choice of the
special vertex $v_1$ or the point $x \in \R^3$.  Assuming the integral
does converge, we can use the Kirillov trace formula
\eqref{E:KirillovKernel} to reexpress it as:
\begin{equation}
\label{E:euclideanEvaluation2}
   \euclideanIntegral{a} = \int_{\!\!\prod\limits_{v \in V'}\!\!\R^3} \prod_{e \in E}
            \left[ \euclideanKernel{a_e}{|x_{s(e)}-x_{t(e)}|} \right]
            \; \prod_{v\in V'} \frac{dx_v}{2\pi^2} .
\end{equation}

When the evaluation of a degenerate spin network converges, it always
obeys a very simple scaling law as we multiply all the edge labels
$a_e$ by the same constant $\smult$.  Using the scaling property
of the degenerate kernel noted in equation~\eqref{E:euclideanKernelScaling},
we find that
\begin{equation}
\label{E:euclideanEvaluationScaling}
    \euclideanIntegral{\smult a} = 
\smult^{|E| - 3(|V| - 1)}\; \euclideanIntegral{a} ,
\end{equation}
where $|E|$ is the number of edges in the underlying graph and
$|V|$ is the number of vertices.

We have already argued that the asymptotics of the Riemannian $10j$
symbols are governed by the corresponding degenerate spin network.  We
can generalize this argument as follows.  Fix a connected graph.  If
we label each edge $e$ by a positive integer $a_e$ --- or equivalently
a spin $j_e$ with $a_e = 2j_e + 1$ --- we obtain a Riemannian spin 
network, whose evaluation we define by:
\begin{equation}
\label{E:riemannianEvaluation}
   \riemannianIntegral{a} = \int_{\!\!\prod\limits_{v \in V'}\!\!S^3} \prod_{e \in E}
             \riemannianKernel{a_e}{d(x_{s(e)},x_{t(e)})} 
            \; \prod_{v\in V'} \frac{dx_v}{2\pi^2} .
\end{equation}
Here $d(x,y)$ is the distance between points $x,y$ in the unit
3-sphere in $\R^4$ as measured by the induced Riemannian metric.  This
formula is equivalent to the standard integral
formula~\cite{Evaluation}, except that we have omitted the usual signs
in order to simplify the relationship to degenerate spin networks.  
Fixing a small open ball $U$ around some point 
$(x,\dots,x) \in \prod_{v \in V'} S^3$ we define the `degenerate
contribution' to this integral to be:
\begin{equation}
\label{E:riemannianDegenerateEvaluation}
   \riemannianIntegralDeg{a} = 2^{|V|-1} \int_{U} \prod_{e \in E}
             \riemannianKernel{a_e}{d(x_{s(e)},x_{t(e)})} 
            \; \prod_{v\in V'} \frac{dx_v}{2\pi^2} .
\end{equation}
Just as we included a factor of $16$ in 
equation~\eqref{E:degenerateContribution},
here we include a factor of $2^{|V|-1}$ to take into account
the contribution of anti-parallel degenerate points; as before
there is no cancellation between these if the spins labelling
edges incident to each vertex sum to an integer, as they must for
the spin network to have a nonzero value.  Using the same 
nonrigorous argument as in Section \ref{S:Degenerate}, we see that as 
$\smult \to \infty$, 
\begin{equation}
\label{E:riemannianDegenerateAsymptotics}
\begin{split}
   \riemannianIntegralDeg{\smult a} 
&\sim   2^{|V|-1} \int_{U} \prod_{e \in E}
        \euclideanKernel{\smult a_e}{|x_{s(e)}-x_{t(e)}|} 
         \; \prod_{v\in V'}  \frac{dx_v}{2\pi^2} \\
&=   2^{|V|-1}\; \smult^{|E| - 3(|V| - 1)} \int_{\smult U} \prod_{e \in E}
         \euclideanKernel{a_e}{|y_{s(e)}-y_{t(e)}|} 
         \; \prod_{v\in V'} \frac{dy_v}{2\pi^2} \\
&\sim  2^{|V|-1} \; \smult^{|E| - 3(|V| - 1)} \; \euclideanIntegral{a}  \\
&=   2^{|V|-1} \; \euclideanIntegral{\smult a}  , 
\end{split}
\end{equation}
where $U$ now denotes an open ball around the origin of 
$\prod_{v \in V'} \R^3$, and we made the change of variables 
$y_e = \smult x_e$.   

In short, this argument suggests that the 
asymptotics of the degenerate contribution to the value
of a Riemannian spin network are proportional to those of the 
corresponding degenerate spin network:
\[     \riemannianIntegralDeg{\smult a} \sim 
2^{|V|-1} \; \euclideanIntegral{\smult a} , \] 
and we know the latter are very simple:
\[     \euclideanIntegral{\smult a}  = 
\smult^{|E| - 3(|V| - 1)} \; \euclideanIntegral{a}  .\]
This is particularly interesting when we also have
\[     \riemannianIntegral{\smult a} \sim 
\riemannianIntegralDeg{\smult a}  ,\]
because then we can compute the asymptotics of a Riemannian spin network
by evaluating a degenerate spin network:
\[   \riemannianIntegral{\smult a} \sim 
2^{|V|-1} \; \smult^{|E| - 3(|V| - 1)} \; \euclideanIntegral{a}.  \] 
When can we expect this to occur?  Clearly we should at least demand that the
degenerate contribution outweigh the contribution of stationary phase
points.  A simple power-counting argument as in Section~\ref{S:Stationary}
suggests that the contribution of stationary phase points is of order:
\begin{equation}
\label{E:riemannianStationaryAsymptotics}
\riemannianIntegralStat{\smult a} = 
\begin{cases} 
	O(\smult^{-\frac{3}{2}|V| + 3}) & |V| > 2 \\
	O(\smult^{-\frac{1}{2}})        & |V| = 2 \\
	O(\smult)                       & |V| = 1 ,
\end{cases}
\end{equation}
where the graphs with one or two vertices are different because there is
less need for `gauge-fixing'.  Comparing these asymptotics to those
of the degenerate contribution, we can formulate the following:

\begin{conj} Given a connected graph with more than
two vertices and $|E| > \frac{3}{2}|V|$, or two vertices and
$|E| > 2$, or one vertex and $|E| > 1$, as $\smult \to \infty$ 
we have
\[
 \riemannianIntegral{\smult a} \sim 
2^{|V|-1}\; \smult^{|E| - 3(|V| - 1)}\; \euclideanIntegral{a}  
\]
as long as the integral defining $\euclideanIntegral{a}$ converges
and the spins $j_e$ labelling the edges incident to each vertex are
admissible.
\end{conj}

\noindent 
Here we say the spins labelling the edges incident to some vertex are
`admissible' if they sum to an integer and each is less than or equal
to the sum of the rest.   We do not yet have general criteria for when 
the integrals associated to Euclidean spin networks converge, and as we
shall see, the relevant theorems are bound to be a bit different than in
the Lorentzian case~\cite{BBL}.

The simplest test of this conjecture is the `theta network', with two
vertices joined by three edges, labelled by positive integers $a$,
$b$, and $c$.  When the corresponding spins are admissible, 
the Riemannian theta network evaluates to:
\begin{equation}
\label{E:riemannianTheta}
\left(\thetagraph{a}{b}{c}\right)^R = 1 .
\end{equation}
The degenerate theta network can also be explicitly evaluated;
assuming without loss of generality that $a \leq b \leq c$:
\begin{equation}
\label{E:euclideanTheta}
\left(\thetagraph{a}{b}{c}\right)^D =
\begin{cases}
0                   & c > a+b \\
\frac{1}{4}         & c = a+b \\
\frac{1}{2}         & c < a+b .
\end{cases} 
\end{equation}
Since the Riemannian network's spins are admissible, the third
inequality must hold for the corresponding areas $a,b,c$.  Thus in this
case the conjecture gives an \emph{exact} formula for the Riemannian
spin network.

The next simplest case is the `4$j$ symbol': the spin network with 
two vertices joined by four edges, labelled by positive integers
$a$, $b$, $c$ and $d$.  
Without loss of generality let us assume $a \leq b \leq c \leq d$.  
As noted in a previous paper in this series~\cite{Riemannian},
the Riemannian 4$j$ symbol counts the dimension of a space
of $\SU(2)$ intertwiners.  Using this it follows that:
\begin{equation}
\label{E:riemannianFourta}
\left(\fourtheta{a}{b}{c}{d}\right)^R = 
\begin{cases}
0                    & b+c \leq d-a \\
\frac{1}{2}(a+b+c-d) & d-a \leq b+c < d+a \\
a                    & d+a \leq b+c .
\end{cases}
\end{equation}
The corresponding degenerate spin network evaluates to:
\begin{equation}
\label{E:euclideanFourta}
\left(\fourtheta{a}{b}{c}{d}\right)^D = 
\begin{cases}
0                    & b+c \leq d-a \\
\frac{1}{4}(a+b+c-d) & d-a \leq b+c < d+a \\
\frac{1}{2} a        & d+a \leq b+c , 
\end{cases}
\end{equation}
so the conjecture is again exact.

An interesting check on our hypotheses is the tetrahedral spin
network.  This has four vertices and $|E| = \frac{3}{2}|V|$, so the
hypotheses of Conjecture 2 do not apply: we expect the stationary
phase contribution to the Riemannian tetrahedral network to be
comparable to the degenerate contribution.   The degenerate tetrahedral
network evaluates to:
\begin{equation}
\label{E:euclideanTet}
\left(\TetJ{a}{b}{c}{d}{e}{f}\right)^D = \frac{1}{24 \pi \V(a,b,c,d,e,f)} ,
\end{equation}
where $\V(a,b,c,d,e,f)$ is defined as the volume of the tetrahedron dual
to the tetrahedral network.  Each triangle in this dual tetrahedron
corresponds to a vertex of the tetrahedral network, and the
three sides of the triangle have lengths equal to the labels on 
the three edges incident to the network vertex:
\begin{equation}
\label{E:tetVolume}
\V(a,b,c,d,e,f) = \text{the volume of}\TetJ{f}{d}{e}{b}{c}{a} .
\end{equation}
On the other hand, the Riemannian tetrahedral
network evaluates to the square of the $\SU(2)$ tetrahedral network,
the basic building-block of the Ponzano--Regge model.  Thanks to
the calculation of Ponzano and Regge \cite{PR}, later made rigorous
by Roberts \cite{Roberts}, this means that:
\begin{equation}
\label{E:riemannianTet}
\begin{split}
\left(\TetJ{a}{b}{c}{d}{e}{f}\right)^R &\sim
\frac{\cos^2 (S + \frac{\pi}{4})}
{12 \pi 
\V(\frac{a}{2},\frac{b}{2},\frac{c}{2},\frac{d}{2},\frac{e}{2},\frac{f}{2})} \\
& =
\frac{1 + \cos 2 (S + \frac{\pi}{4})}
{24 \pi 
\V(\frac{a}{2},\frac{b}{2},\frac{c}{2},\frac{d}{2},\frac{e}{2},\frac{f}{2})}.
\end{split}
\end{equation}
Here we are dealing with a dual tetrahedron whose edge lengths are
$\frac{1}{2}a_{kl}$, where $a_{kl}$ ranges over $a,b,c,d,e,f$ as 
$1 \le k < l \le 4$.  This tetrahedron has Regge action
\[  S = 
\sum_{1 \le k < l \le 4}  \textstyle{\frac{1}{2}} a_{kl} \theta_{kl} ,\]
where $\theta_{kl}$ are the corresponding dihedral angles.  
Thus it appears that the asymptotics of the Riemannian
tetrahedral network are a sum of two parts: a part equal to 8 times the
degenerate tetrahedral network, and an oscillatory part coming from 
the stationary phase points.

When the edge lengths of the dual tetrahedron are such that it cannot exist in
Euclidean space, the degenerate tetrahedral network evaluates to zero, and
the Riemannian network obeys different asymptotics in which its evaluation
exponentially decays with increasing spin.

\section{Lorentzian Spin Networks}
\label{S:Lorentzian}

Now let us turn to Lorentzian spin networks~\cite{BCL,BBL}.  Each
positive real number determines a unitary irreducible representation
of the connnected Lorentz group $\SO_0(3,1)$ corresponding to an eigenspace 
of the Laplacian on $H^3$; however, we prefer to describe the evaluation
using an integral formula.  Fixing a connected graph 
with vertex set $V$ and edge set $E$, and labelling
each edge $e$ by a positive real number $a_e$, we obtain a so-called
`Lorentzian spin network', whose evaluation is given by:
\begin{equation}
\label{E:LorentzianEvaluation}
   \lorentzianIntegral{a} = \int_{\!\!\prod\limits_{v \in V'}\!\!H^3} \prod_{e \in E}
             \lorentzianKernel{a_e}{d(x_{s(e)},x_{t(e)})} 
            \; \prod_{v\in V'} \frac{dx_v}{2\pi^2} .
\end{equation}
As with degenerate spin networks, we have chosen a vertex $v_1$,
set $V' = V - \{ v_1 \}$, and let $x_{v_1}$ be any fixed point in $H^{3}$.
Here $H^3$ is hyperbolic 3-space, i.e., the 
submanifold of Minkowski spacetime given by:
\[  H^3 = \{t^2 - x^2 - y^2 - z^2 = 1 , \; t > 0 \}  \]
with its induced Riemannian metric.  We define the Lorentzian kernel
$K^L$ by:
\begin{equation}
\label{E:lorentzianKernel}
\lorentzianKernel{a}{\phi} := \frac{\sin a\phi}{\sinh \phi}.
\end{equation}
We warn the reader that this convention differs from that of most
previous papers~\cite{Positivity,BCL,BBL}, which include a factor
of $a$ in the denominator.  Including that factor would divide
any Lorentzian spin network by the product of its edge labels, so
for example, it would divide the asymptotics of the Lorentzian 
$10j$ symbols as defined here by a factor of $\smult^{10}$. 

The same line of argument by which we arrived at our conjecture
concerning asymptotics of Riemannian spin networks applies to
Lorentzian ones.  The most important difference is that no
factor of $2^{|V|-1}$ appears, since there are no `antipodal points'
in hyperbolic space.  We have not investigated criteria for the existence
of stationary phase points, but where they are present
their exponents will be the same as in the Riemannian
case, leading us to make:

\begin{conj} Given a connected graph with more than
two vertices and $|E| > \frac{3}{2}|V|$, or two vertices and
$|E| > 2$, or one vertex and $|E| > 1$, as $\smult \to \infty$ we have
\[
   \lorentzianIntegral{\smult a} \sim \smult^{|E| - 3(|V| - 1)}\; 
   \euclideanIntegral{a}  
\]
as long as the integral defining $\euclideanIntegral{a}$ converges
and the positive numbers $a_e$ labelling edges incident to each
vertex are admissible.  
\end{conj}

\noindent Here we say the positive numbers labelling the edges incident
to some vertex are `admissible' if each is strictly less than 
the sum of the rest.

Again the simplest test of this conjecture is the Lorentzian
theta network.  Translating their result into our notation, a 
calculation of Barrett and Crane~\cite{BCL} shows that for any
$a,b,c > 0$, 
\begin{equation*}
%\label{E:lorentzianTheta}
\left(\thetagraph{a}{b}{c}\right)^L = 
\frac{1}{4} 
\left[f(-a + b + c) + f(a - b + c) + f(a + b - c) - f(a + b + c) \right] ,
\end{equation*}
where 
\[     f(k) = \tanh(\frac{\pi}{2} k)  .\]
As the conjecture predicts, the asymptotics of this match those of the
degenerate theta network, which are given by:
\begin{equation*}
%\label{E:euclideanTheta2}
\left(\thetagraph{a}{b}{c}\right)^D = 
\frac{1}{4} 
\left[\sign(-a+b+c) + \sign(a-b+c) + \sign(a+b-c) - \sign(a+b+c)\right] .
\end{equation*}
It is worth noting that while the integral for the Lorentzian theta
network converges even when we take the absolute value of the integrand, 
this fails for the degenerate theta network.  This makes it more 
challenging to find criteria for convergence of degenerate spin networks,
since we cannot simply mimic the theory that applies in the 
Lorentzian case~\cite{BBL}.

Barrett and Crane also worked out the Lorentzian $4j$ symbols, obtaining:
% With this joined up, the label doesn't fit.  But we never refer to it,
% so let's just omit it.
\begin{equation*}
%\label{E:lorentzianFourta}
\begin{split} 
\left(\!\fourtheta{a}{b}{c}{d}\!\right)^{\!L} \!= 
\frac{1}{4} [g(-&a+b+c+d)+g(a-b+c+d)+g(a+b-c+d)+g(a+b+c-d)\\
             -g(&a+b-c-d)-g(a-b+c-d)-g(a-b-c+d)-g(a+b+c+d) ] ,
\end{split}
\end{equation*}
where
\[ g(k)=\frac{k}{2} \coth( \frac{\pi}{2} k).\]
From equation~\eqref{E:euclideanFourta} one can show:
\[   \left(\fourtheta{a}{b}{c}{d}\right)^D = \]
\[
\begin{split} 
\frac{1}{4} [h(-&a+b+c+d)+h(a-b+c+d)+h(a+b-c+d)+h(a+b+c-d)\\
-h(&a+b-c-d)-h(a-b+c-d)-h(a-b-c+d)-h(a+b+c+d) ] ,
\end{split}
\] 
where 
\[ h(k)=\frac{k}{2} \sign(k) .\]
Thus the conjecture is also confirmed in this case.

Next, consider the tetrahedral spin network.  As in the Riemannian case,
Conjecture 3 does not apply, but we can still predict the
asymptotic behaviour of the contribution of the fully degenerate point:
\[
\begin{split}
 \left(
 \TetJ{\smult}{\smult}{\smult}{\smult}{\smult}{\smult}
 \right)^L &\sim
  \frac{1}{24 \pi \V(\smult,\smult,\smult,\smult,\smult,\smult)} + SP\\
 &= \frac{\sqrt{2}}{4 \pi} \smult^{-3} + SP,
\end{split}
\]
where `$SP$' represents the contribution from stationary phase points, if any.
Below is a log-log plot comparing this prediction to numerical data.
The horizontal axis in this graph represents $\smult$, while the vertical
axis represents the value of the tetrahedral spin network.
{
\center
\includegraphics{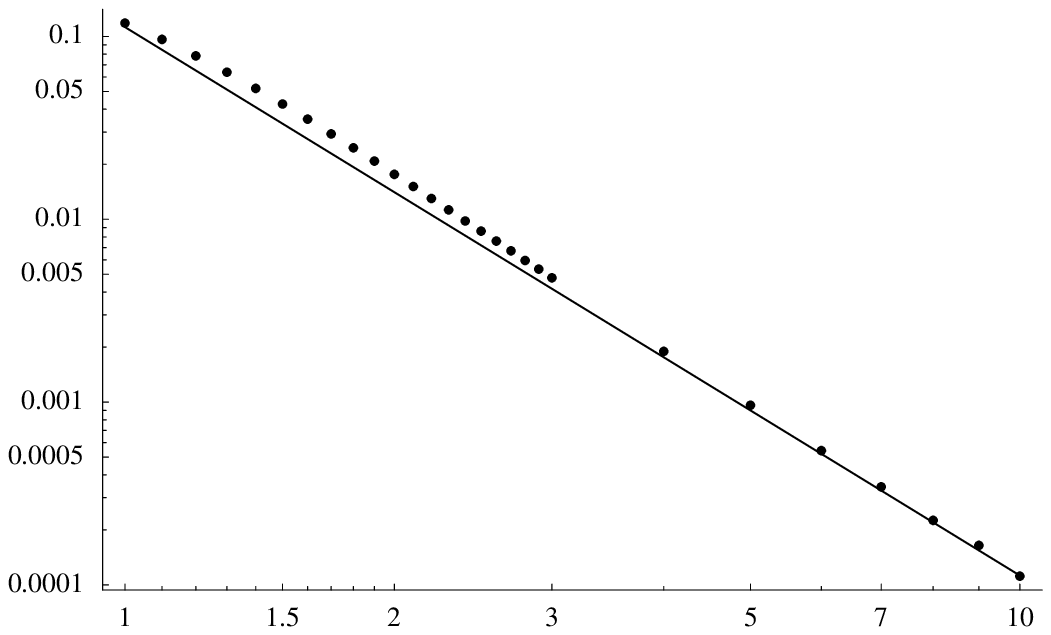}
}

The most interesting test of Conjecture 3 is the $10j$ symbol.  If the
conjecture is true, the Lorentzian $10j$ symbol should be asymptotic to
the degenerate $10j$ symbol, and therefore asymptotic to $1/16$ times
the Riemannian $10j$ symbol.  Since the Riemannian $10j$ symbol is 
positive \cite{Positivity}, this in turn would imply that the Lorentzian
and degenerate $10j$ symbols are positive in the $\smult \to \infty$
limit.  

It is difficult to compute the Lorentzian $10j$ symbol, but we have
numerically checked the conjecture in the special case where
all the edges are labelled by the same number $\smult$.  In this case 
the conjecture says that:

\[
\begin{split}
 \left( \; \TenL \;\; \right)^L &:=
  \int_{(H^3)^4} 
  \prod_{k<l} \lorentzianKernel{\smult}{\phi_{kl}}
  ~\frac{dx_2}{2\pi^2} \cdots \frac{dx_5}{2\pi^2}  \\
  & \\
 &\sim  .1706 \, \smult^{-2} ,
\end{split}
\]
where of course the constant is not exact.
Below we show a log-log plot comparing this prediction to data points
obtained by computing the Lorentzian $10j$ symbols numerically.  The
horizontal axis represents $\smult$, while the vertical
axis represents the value of the $10j$ symbol.

%\vfill \eject
{
\center
\includegraphics{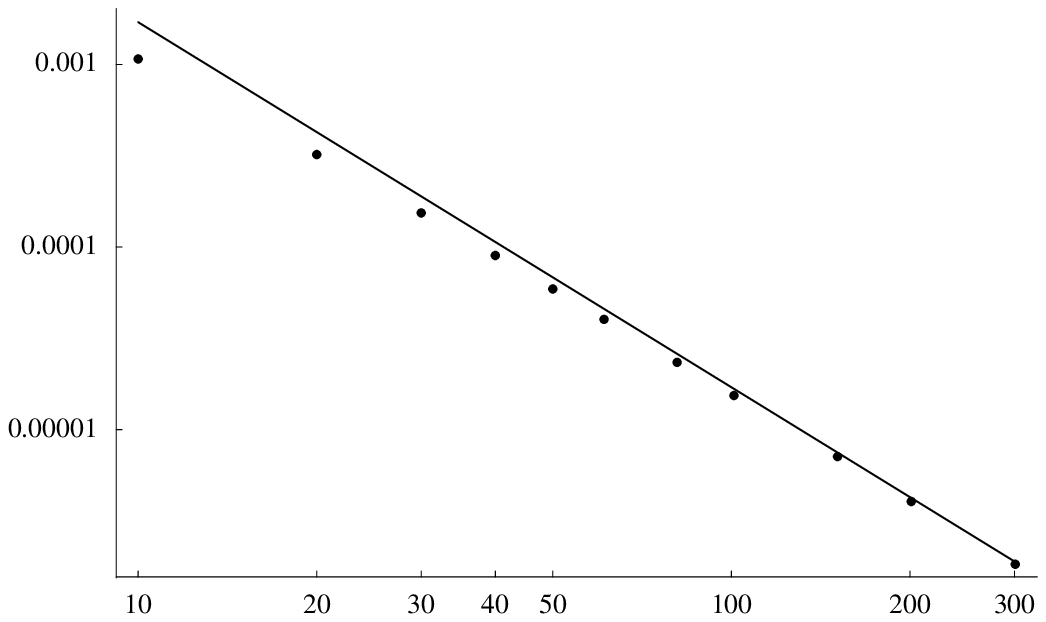}
}

\noindent 
Here we computed the Lorentzian $10j$ symbols by applying the VEGAS
algorithm to evaluate the integral above
using polar coordinates on $H^3$.  The integral was reduced from 12 to
9 dimensions by exploiting the $\SO_0(3,1)$ invariance; the infinite
domain was made compact by replacing the radial coordinate for each
point, $r_i$, with a new variable $t_i = r_i / (1+r_i)$; and the
domain was further reduced by exploiting a 24-fold symmetry present in
the regular case.  The large dimension and oscillatory nature of this
integral make these calculations extremely computationally intensive.  

The graph below shows the same Lorentzian $10j$ symbols multiplied by 
$\smult^2$, plotted on a linear scale to give a clearer picture of the rate 
of convergence towards the asymptote.  The error bars are three times the 
standard deviation computed by the VEGAS algorithm.  The coordinate 
system we used yielded the lowest standard deviations of several we 
tried, but there is no reason to believe that the estimates of the 
integral are drawn from a normal distribution, so this data should be 
treated with caution.

{
\center
\includegraphics{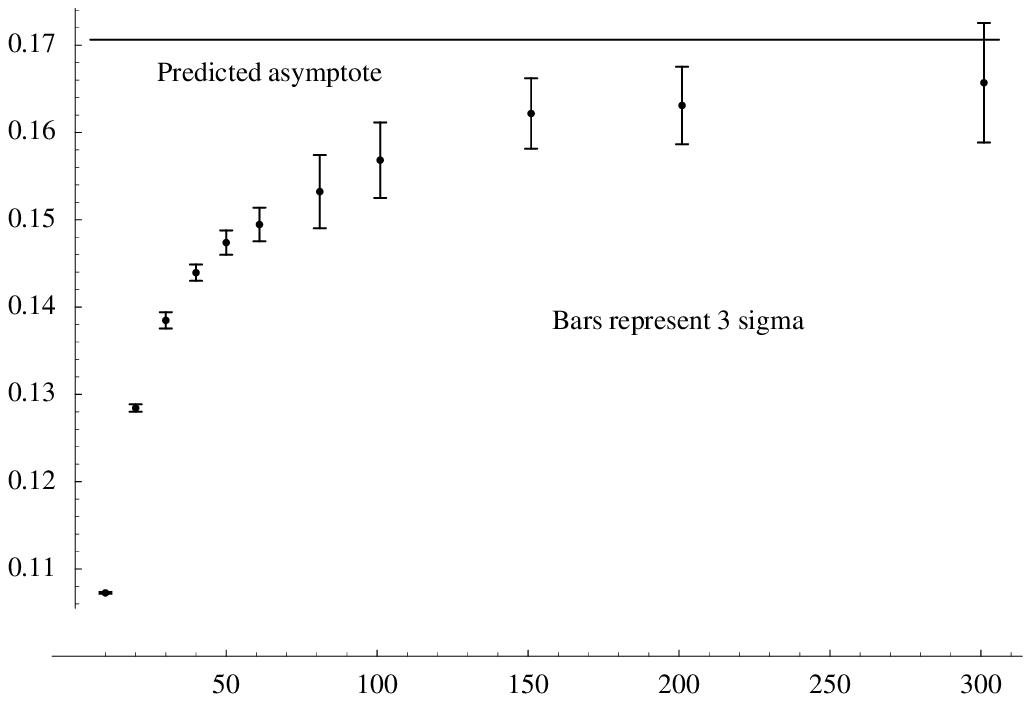}
}

\section{Conclusions}
\label{S:Conclusions}

It appears that the asymptotics of both Riemannian and Lorentzian
$10j$ symbols can be computed in terms of degenerate spin networks.
For the mathematician, this claim still requires proof.  Indeed, there
is even an issue of rigor concerning our argument that degenerate
spin networks asymptotically describe the
contribution of fully degenerate points, since we have not proved
that the limit
\[   \lim_{\smult \to \infty} \int_{\smult U} \prod_{e \in E}
         \euclideanKernel{a_e}{|y_{s(e)}-y_{t(e)}|} 
         \; \prod_{v\in V'} \frac{dy_v}{2\pi^2} \]
exists.  The ambitious mathematician could also try to prove more
general versions of our conjectures, in which all the areas $a_e$
approach infinity, but not in fixed proportion to one another.

For the physicist, however, a more pressing question is: \emph{what
do these results imply for the physics of the Barrett--Crane model?}
On the one hand, it is unsettling that the simple asymptotic behavior
of the Ponzano--Regge model is not found here.  The way degenerate
geometries govern the asymptotics of the $10j$ symbols raises the
possibility that in the limit of large spins, the Barrett--Crane model
reduces to a theory of degenerate metrics.  However, it is important
to bear in mind that the physics of the Barrett--Crane model may not
be controlled by the large-spin behavior of the $10j$ symbols ---
though certainly this affects the convergence of the partition
function \cite{Riemannian}.  Indeed, there are still major open
questions about the right way to extract physics from spin foam
models, and we hope that our work spurs on research on this subject.

\vskip 1em

\subsection*{Note Added in Proof}

After we submitted this article for publication, preprints of two
forthcoming papers appeared, one by John Barrett and Chris Steele
\cite{BS} and another by Laurent Freidel and David Louapre \cite{FL},
both of which deal with the asymptotics of $10j$ symbols by somewhat
different methods than our own, and both of which obtain results in
agreement with our conjectures.

\subsection*{Acknowledgements}

We thank John Barrett and Chris Steele for helpful conversations, and
note that we corrected two errors in an earlier draft of this paper
(an excess factor of 2 in equation~\eqref{E:statPhaseResult}
and a false conjecture about stationary phase points in
Lorentzian spin network evaluations) after reading
a preprint of their forthcoming paper on the asymptotics of $10j$ symbols.

We also thank SHARCNet for providing
the supercomputer at the University of Western Ontario on which we did
some of our computations.

\revisions { v1.0 28 Sep 2001 First draft (GE) \\ v1.1 30 Sep 2001
Clarified multiplication of spins vs. areas (GE) \\ v1.2 7 Oct 2001
Added formulas for degenerate theta, four-edged theta, and tet (GE) \\
v1.3 28 Jan 2002 Added two more data sets to asymptotics plot (GE) \\
v1.4 30 Jan 2002 Used different symbols for different data sets in plot.
 Noted that we haven't proved convergence of
\eqref{E:coefficientIntegral}. (GE) \\ v1.5 4 Feb 2002 Second plot to
show exponential decay in some cases. (GE) \\
v1.6 17 Jul 2002 Polished the wording, added more discussion of
degenerate 4-simplices, degenerate spin networks, and 
relevance to quantum gravity --- not done yet. (JB) \\
v1.7 23 Jul 2002 Improved graphs, more on degenerate spin networks, 
wording polished --- a few remaining issues marked by ???. (JB) \\
v1.8 24 Jul 2002 Incorporated corrections from Dan and Greg, split
off section on Lorentzian spin networks, more on degenerate spin
networks --- a few remaining issues marked by ???. (JB) \\
v1.9 29 Jul 2002 Lots of errors corrected by Greg and Dan,
new stuff on Lorentzian tet net, error bars for Lorentzian 10j, 
rescaling areas instead of spins.  Help from Greg requested at
points marked by ???. (JB) \\
v1.10 30 Jul 2002 Corrected some minor typos; rewrote section leading
up to 5-dimensional integral for degenerate $10j$; redrew diagrams and
rewrote accompanying text to reflect area scaling of Riemannian $10j$,
removing examples of exponential decay in spin scaling;
slight corrections to Riemannian-degenerate theta/fourta/tet comparisons 
(GE) \\
v1.11 31 Jul 2002 Put back plots of spin scaling, added some commentary on
spin vs area scaling (GE) \\
v1.12 2 Aug 2002 Corrected description of degenerate 4-simplices,
expanded discussion of spin vs area scaling, some smaller changes. (JB) \\
v2 5 Aug 2002 Version submitted to gr-qc. (JB) \\
v2.1 28 Sep 2002 Addressed factor of 2 mistake in coefficient for stationary
phase term in Riemannian $10j$, and removed false claims of non-existence
of stationary phase points in Lorentzian integrals (GE) \\
v2.2 18 Oct 2002 Mentioned preprints of Barrett and Steele, and Freidel and
Louapre papers, revised acknowledgements (GE) \\
v2.3 20 Oct 2002 Revised mention of preprints, adjusted page breaks (JDC) \\
}


\begin{thebibliography}{99}

\bibitem{PR} G.\ Ponzano and T.\ Regge, Semiclassical limit of
Racah coefficients, in {\sl Spectroscopic and Group Theoretical
Methods in Physics,} ed.\ F.\ Bloch, North--Holland, New York, 1968.

\bibitem{Roberts} J.\ Roberts, Classical $6j$-symbols and the
tetrahedron, {\sl Geom.\ Top.\ }{\bf 3} (1999), 21--66.
Also available as math-ph/9812013.

\bibitem{Roberts2} J.\ Roberts, Asymptotics and 6j-symbols, available 
as math.QA/0201177.

\bibitem{BC}
J.\ W.\ Barrett and L.\ Crane,
Relativistic spin networks and quantum gravity,
{\sl Jour.\ Math.\ Phys.\ } \textbf{39} (1998), 3296--3302.
Also available as gr-qc/9709028.

\bibitem{Positivity}
J.\ C.\ Baez and J.\ D.\ Christensen,
Positivity of spin foam amplitudes, {\sl Class.\ Quantum Grav.\ } 
{\bf 19} (2002), 2291--2306.
Corrected version available as gr-qc/0110044.

\bibitem{Riemannian}
J.\ C.\ Baez, J.\ D.\ Christensen, T.\ R.\ Halford and D.\ C.\ Tsang,
Spin foam models of Riemannian quantum gravity,
to appear in Classical and Quantum Gravity. 
Available as gr-qc/0202017.

\bibitem{Barrett}
J.\ W.\ Barrett,
The classical evaluation of relativistic spin networks,
{\sl Adv.\ Theor.\ Math.\ Phys.\ }\textbf{2} (1998), 593-600.
Also available as math.QA/9803063.

\bibitem{BW}
J.\ W.\ Barrett and R.\ M.\ Williams,
The asymptotics of an amplitude for the 4-simplex,
{\sl Adv.\ Theor.\ Math.\ Phys.\ }\textbf{3} (1999), 209--215.
Also available as gr-qc/9809032.

\bibitem{Algorithm}
J.\ D.\ Christensen and G.\ Egan,
An efficient algorithm for the Riemannian $10j$ symbols,
{\sl Class.\ Quantum Grav.\ }{\bf 19} (2002), 1185--1194.
Also available as gr-qc/0110045.

\bibitem{Ashtekar} A.\ Ashtekar, Quantum mechanics of geometry, 
available as gr-qc/9901023.

\bibitem{APB}  A.\ Alekseev, A.\ P.\ Polychronakos and M. Smedb\"ack,
On area and entropy of a black hole, available as hep-th/0004036.

\bibitem{BCL}
J.\ W.\ Barrett and L.\ Crane,
A Lorentzian signature model for quantum general relativity,
{\sl Class.\ Quantum Grav.\ }\textbf{17} (2000), 3101--3118.
Also available as gr-qc/9904025.

\bibitem{FK} L.\ Freidel and K.\ Krasnov, 
Simple spin networks as Feynman graphs, 
{\sl J.\ Math.\ Phys.\ }{\bf 41} (2000), 1681--1690.
Also available as hep-th/9903192.

\bibitem{StatPhase}
A.\ Erdelyi,
{\sl Asymptotic Expansions},
Dover Publications, New York, 1954.

\bibitem{Bang}
H.\ Mowaffaq, 
A vector proof of a theorem of Bang, 
{\sl Amer.\ Math.\ Monthly} {\bf 108} (2001), 562--564. 

\bibitem{Tetrahedron}
J.\ C.\ Baez and J.\ W.\ Barrett, 
The quantum tetrahedron in dimensions 3 and 4, 
{\sl Adv.\ Theor.\ Math.\ Phys.\ }{\bf 3} (1999), 815--850. 
Also available as gr-qc/9903060.

\bibitem{Vegas}
G.\ P.\ Lepage,  A new algorithm for adaptive multidimensional integration,
{\sl Journal of Computational Physics} \textbf{27} (1978), 192--203.

\bibitem{Foam} 
J.\ C.\ Baez, 
An introduction to spin foam models of $BF$ theory and quantum
gravity, in {\sl Geometry and Quantum Physics}, eds.\ Helmut 
Gausterer and Harald Grosse, Springer Lecture Notes in 
Physics {\bf 543}, Berlin, 2000.  Also available as gr-qc/9905087.

\bibitem{BBL}
J.\ C.\ Baez and J.\ W.\ Barrett, 
Integrability for relativistic spin networks, 
{\sl Class.\ Quant.\ Grav.\ }{\bf 18} (2001), 4683--4700.
Also available as gr-qc/0101107.

\bibitem{Evaluation}
J.\ W.\ Barrett, 
The classical evaluation of relativistic spin networks, 
{\sl Adv.\ Theor.\ Math.\ Phys.\ }{\bf 2} (1998), 593--600.
Also available as math.QA/9803063.

\bibitem{BS} J.\ W.\ Barrett and C.\ M.\ Steele, Asymptotics of
relativistic spin networks, available as gr-qc/0209023.

\bibitem{FL} L.\ Freidel and D.\ Louapre, Asymptotics of 6j and 10j
symbols, available as hep-th/0209134.

\end{thebibliography}
\end{document}